\documentclass[reprint]{JASA}
\usepackage{multirow}
\usepackage{ifthen}
\newboolean{color}
\usepackage{dirtytalk}

\begin{document}
\setboolean{color}{true}

\title[Sound emergence and WT Noise]{Sound emergence as a predictor of short-term annoyance from wind turbine noise}


\author{Elise Ruaud}
\author{Guillaume Dutilleux}


\affiliation{Acoustics Group, Department of Electronic Systems,  NTNU, Trondheim, 7491,  Norway }


\altaffiliation{}


\email{guillaume.dutilleux@ntnu.no}




\begin{abstract}
While sound emergence is used in several countries to regulate wind energy development, there is no published evidence that it is a relevant noise descriptor for this purpose. In the present work, we carried out two listening tests to evaluate the merits of sound emergence. Three definitions of sound emergence were considered: the one in ISO 1996-1, sound emergence \emph{under audibility condition} $e_{UAC}$, and \emph{spectral emergence} $e_{SP}$. We also considered the \emph{specific to residual ratio} and loudness metrics. In each listening test, the sound stimuli consisted of 48 sound stimuli at 3 A-weighted sound pressure levels $\{30, 40, 50\}$~dB and 4 specific-to-residual ratios $\{-10, -5, 0, +5 \}$~dB. The results lead to the conclusion that short term annoyance is better predicted by the total sound pressure level than by sound emergence, whatever the definition considered for the latter, or than by the specific to residual ratio. Short-term annoyance is slightly better predicted by $e_{UAC}$ than by $e$, while $e$ is a better predictor than $e_{SP}$. The total sound pressure level and the loudness metrics performed similarly. Furthermore, the results provide evidence that sound emergence is a poor predictor of the audibility of wind turbine sounds.

\end{abstract}


\maketitle


\section{Introduction}


Several authors have provided empirical evidence that wind turbine noise can be a source of annoyance \cite{Pedersen:2004aa,Pedersen:2009aa,Janssen:2011aa, Schaffer:2016ww, Pohl:2018aa, Michaud:2016tj}, even though it is well known that noise explains only a small part of the variance of annoyance due to noise-generating activities in general \cite{Stallen:1999aa} and wind energy in particular \cite{Pohl:2018aa,Haac:2019uv}. Recent research even asserts that sound pressure level is not the dominant predictor of noise annoyance and highlights the role of project participation and visual disturbance in noise annoyance among a number of other factors \cite{Haac:2019uv}. Nevertheless, acoustics remains a key factor in the acceptance of wind energy by the communities. Therefore, it is worth considering the relevance of the metric used for setting legal noise limits. Some authors have also found that living in an area with a subjectively expected low background sound pressure level is positively associated with annoyance from wind turbine sounds \cite{Pedersen:2007vp}. Moreover, it has been observed that natural sounds that are rated positively may mask wind turbine sounds to some degree, which may result in reduced annoyance \cite{Bolin:2012aa}. This leads the latter research to the suggestion of setting limit values "in terms of relative levels to the background sound". 


In the particular context of wind turbine noise management, most of the countries rely on various ratings based on either (1) the \emph{total} sound only, or (2) the \emph{specific} sound - or source-attributable sound - and the \emph{residual} sound, \emph{i.e} the remaining sound without any source-attributable sound \cite{Davy:2018aa, Hathaway:2006ub}. But some legislation and guidelines prefer to rely on (3) \emph{sound emergence} $e$. 

To our knowledge only two countries around the world, namely Italy \cite{ministri:1991aa,Italy:1997aa,Italy:2004aa} and France \cite{France:2011aa,France:2020vp}, specify sound emergence limit values. However, the concept of sound emergence appears implicitly in the guidelines issued by the World Bank \cite{WorldBank:2007aa,World-Bank:2015aa} as well. Therefore, sound emergence is likely to be used in many countries where wind farm projects receive funding from the World Bank. These guidelines specify a maximum increase in "background levels of 3 dB at the nearest receiver location". The 3 dB value is always combined with limit values expressed as $L_{eq}$ and applies beyond the property boundaries of the noisy facilities. The current compliance criterion is that emergence must be lower than 3 dB for night-time and 5 dB for day-time in Italy and France \cite{Italy:1997aa,France:2011aa,France:2020vp}. The field assessment of the compliance of a wind farm with such low limit values is not devoid of challenges \cite{Dutilleux:2019aa}. The main issue, however, is the psycho-acoustic foundation of sound emergence as a noise descriptor and as a predictor of noise annoyance, if the point is to protect the communities from annoyance due to wind turbine noise.

Dose-response curves have been developed for A-weighted sound pressure levels \cite{Pedersen:2004aa, Janssen:2011aa, Hongisto:2017up} and the merits of C-weighted sound pressure levels with respect to A-weighted ones have also been assessed \cite{Bolin:2014aa}. Yet, it appears that there is no published evidence supporting the use of standard sound emergence as an alternative to sound pressure level in the context of annoyance due to wind turbine sounds \cite{Dutilleux:2019aa}. 

A first paper \cite{Viollon:2004aa} deals with a listening test based on 24 synthetic sounds that were built from three different residual sounds (quiet at $L_{Aeq}=34$~dB, natural at $40$~dB and road traffic at 50 dB) and four different specific industrial sounds including a wind turbine sound at two sound emergences ($e=3$ and $e=5$~dB). It is concluded in \cite{Viollon:2004aa} that the degree of short-term annoyance is source-dependent at constant $e$ and that $e=0$ dB does not imply that the specific sound cannot be detected. Moreover, these authors state that "... in our experiment, the French legal criterion [sound emergence] for assessing noise impact of industrial sources was not appropriate and did not characterize the perceived noise annoyance." In \cite{Viollon:2004aa}, the design of the listening test prevents from evaluating annoyance at various sound levels \emph{ceteris paribus} because the type of residual sound used is sound-pressure-level-dependent. 

Another research already mentioned investigated the reduction in annoyance - or annoyance masking -  and loudness masking of wind turbine sounds from two different turbines when they were mixed with positively rated natural sounds \cite{Bolin:2012aa}. The four natural sounds considered are birdsong, deciduous vegetation sounds, a brook and breaking sea waves. The A-weighted sound pressure level of the stimuli ranged between 26 and 59 dB and the signal-to-noise ratios between -15 and +5 dB. This laboratory study did not deal explicitly with sound emergence in the standard sense but related annoyance to the signal-to-noise ratio. The authors report clear masking effects when both the signal-to-noise ratio is negative and the masking sound features small temporal variations. In our perspective, the two limitations here are, first, the use of very short stimuli (4 seconds) and, second, the focus on positively rated sounds. Using positively rated sounds seems an interesting perspective. In general, however, the background sound is not under our control and the background sound may contain unwanted components. Moreover temporal variations are rather likely to occur in practice and this may compromise masking effects.

More recently, a listening test named $T0$ for later reference was carried out to evaluate the relative merits of sound pressure level and sound emergence as predictors of annoyance from wind turbine noise \cite{Dutilleux:2020aa}. The second author of the present paper is a co-author of \cite{Dutilleux:2020aa}. Three wind turbine sounds, or specific sounds, $\{S0, S1, S2\}$ and one residual sound $R0$ were collected and mixed to prepare a set of synthetic soundscapes. The library of stimuli consisted of 45 30-s sound clips at three different A-weighted sound pressure levels $\{35, 40, 50\;dB\}$ and five different specific-to-residual ratios $\{-10, -5, 0, +5,+10\;dB\}$. The test samples were corrected for atmospheric attenuation over 500~m but did not include any ground effect. Thirty two persons rated the test samples in an acoustically dry room equipped with loudspeakers for playback. The statistical analysis pointed out that short term annoyance was better predicted by A-weighted sound pressure level than by sound emergence. It was also observed that sound emergence performed poorly as a predictor of the audibility of wind turbine sounds, considering that a significant part of the subjects could detect wind turbine sounds even when sound emergence was close to 0 dB. But the need to keep the listening test below a reasonable duration prevented from considering more than one residual sound and more than three different wind turbine sounds in the test samples, so that extending the study was desirable. Moreover, these findings were obtained on a fairly limited number of subjects. 

Considering the limitations of \cite{Dutilleux:2020aa}, the present paper reports about two listening tests carried out in laboratory settings. Its aim is to evaluate  whether sound emergence is a better predictor of short-term annoyance due to aerodynamic wind turbine noise than the total sound pressure level alone. At a receptor location, non-aerodynamic sounds, like tonalities, are not representative of a well-designed and maintained wind turbine \cite{Hau:2016aa}. These additional listening tests offered the opportunity to consider two other residual sounds and three other specific sounds collected specifically. They permitted to introduce ground effect when taking sound propagation into account. Beside the standardized definition of sound emergence, two alternative definitions were considered in the statistical analysis. A second objective was to assess the capacity of sound emergence to predict the audibility of wind turbine sounds. While the present paper deals with a noise indicator that depends on the residual sound, it is not an in-depth investigation of the role of residual sound in annoyance due to wind turbine sounds. Moreover, designing alternative predictors of short term annoyance is beyond the scope or this paper. 

The paper is organized as follows. Section~\ref{sec:defs} provides the definitions of the different metrics to be correlated with short-term annoyance. Section~\ref{sec:methods} presents the methods. It focuses on the collection of audio samples of both specific and residual sound in the field to generate stimuli with controlled total sound pressure level and specific-to-residual ratio, and on the preparation of the listening tests. Section~\ref{sec:results} describes the results obtained, Section~\ref{sec:discuss} discusses (1) the correlation between short-term annoyance and the different quantitative metrics considered, (2) the issue of the relationship between audibility of the specific sound and sound emergence and (3) several methodological aspects. Section~\ref{sec:conclusion} concludes this paper.


\section{Definitions}
\label{sec:defs}

In the remainder of this paper, a first type of sound emergence $e$ is defined as:

\begin{equation}
\label{eq:e}
e=L_{\text{tot}}-L_{\text{res}}.
\end{equation}

Where $L_{\text{tot}}$ is the sound pressure level of the \emph{total sound} and $L_{\text{res}}$ the sound pressure level of the \emph{residual sound}. It is useful to further introduce $L_{\text{spec}}$ for the \emph{specific sound}, \emph{i.e} the sound pressure level attributable to the source under investigation. Here, these four concepts of sound emergence, of total, specific and residual sound are used as defined in the current ISO 1996-1 standard \cite{Dutilleux:2019aa,ISO:2016aa}. Nevertheless, two variants of sound emergence of practical relevance are introduced. The first one is the so-called \emph{emergence under audibility condition} that is non-zero only if the specific sound is audible for / can be detected by a listener in receptor location: 

\begin{equation}
\label{eq:eUAC}
e_{\text{UAC}}=\left\{\begin{array}{ll}
               e&\text{if the specific sound is audible for the subject}\\
               0&\text{otherwise}
        \end{array}
        \right.
\end{equation}

Such a definition is a common interpretation of sound emergence in the French measurement standard that is applicable to neighborhood noise \cite{AFNOR:1996ww}. 

In the following we will also consider the so-called \emph{spectral emergence} $e_{\text{SP}}$ based on octave band sound pressure levels. It is defined as:

\begin{equation}
\begin{split}
    \label{eq:eSP}
    e_{\text{SP}}=\max \left\{L_{\text{tot,octf}}-L_{\text{res,octf}}\right\},\;\\f\in\{125,250,500,1000,2000,4000\}\;Hz
\end{split}
\end{equation}

In France, $e_{\text{SP}}$ used to be a compliance criterion for wind farms before a new legislation entered into force in 2011. Before then, the compliance of wind farms was assessed by reference to the legislation on neighborhood noise \cite{Premier-ministre:2006wx}. To the knowledge of the authors there was no systematic investigation of the impact of the suppression of $e_{\text{SP}}$ from the list of evaluation criteria.

If the residual and the specific sounds are not correlated, it is easy to derive the following relationship between $e$ and the \emph{specific-to-residual ratio} (SRR) \cite{Dutilleux:2019aa}:

\begin{equation}
  e=10\log_{10}\left(1+10^\frac{\text{SRR}}{10}\right).
\end{equation}

where

\begin{equation}
    \label{eq:SRR}
    \text{SRR}=L_{\text{spec}}-L_{\text{res}}
\end{equation}

SRR is used in the legislation on wind turbine noise in a number of countries \cite{Davy:2018aa,Hathaway:2006ub}, where $L_{AF90}$ often serves as a proxy for $L_{\text{spec}}$ and $L_{\text{res}}$.

The loudness level $L_N$ and their variants $L_{N_5}$ and $L_{N_{max}}$, as defined by Zwicker \cite{Fastl:2007aa,ISO:2017ww}, were also considered
 as possible alternatives to $L_{\text{tot}}$, since loudness-based metrics were expected to feature a better correlation with annoyance. 

\section{Methods}
\label{sec:methods}
In this study, the general approach was to simulate outdoor exposure to wind turbine sounds with fine control on $L_{\text{tot}}$ and SRR. The exposure scenario corresponded to a typical distance between a wind farm and a dwelling in a flat landscape. The sound stimuli used in the listening test were obtained by mixing (1) a recording of specific sound, \textit{i.e.} wind turbine sound, in a situation of high signal-to-noise ratio and (2) a recording of residual sound. The naming convention of the specific sounds $Sx$ and residuals sounds $Rx$ is consistent with the labels used in \cite{Dutilleux:2020aa} but the numbering was chosen to avoid overlap, for the sake of the later discussion. 

Since specific sounds were recorded close to the source, it was necessary to pre-process them to account for propagation effects and occasional tonalities before mixing them with the residual sound. The next sections detail the different steps in the preparation of the listening tests, their realization and their statistical analysis. 

Since the relative amplitudes of the collected residual and specific sounds will be manipulated to generate a range of sound pressure levels and a range of sound emergences, calibrating the recording chains is mostly relevant to make sure that the original signals will not undergo too large an amplification that would lead to unrealistically high sound pressure levels and unrealistic sound stimuli. Therefore, orders of magnitude for the sound pressure level are acceptable and non standard calibration procedures can be used.

\subsection{Recording wind turbine sounds}
Wind turbine sounds were collected at two wind farms: Storheia and Bessakerfjellet, both in the commune of {\AA}fjord in Norway. Following the guidelines of the standard for the measurement of the sound power of wind turbines \cite{IEC:2018aa}, the recording campaign was made close to the turbines to ensure a high signal-to-noise ratio. Several recordings were made for different wind turbines and at different angles with the wind direction. Information about the measurement campaigns and the wind farms is provided in \autoref{tab:camp}.

All the acquisitions were made with a 1/2" 200~V class 1 condenser microphone (Brüel and Kjær type 4165) connected to a preamplifier (Norsonic type 336) with a high pass filter at 20 Hz. The preamplifier served as a front end to a digital audio recorder (Sound Devices type 722). The recordings were made with a sampling frequency of 48 kHz and a resolution of 24 bits. A calibration was made before and after each recording session with a class I calibrator (Norsonic type Nor1256).
The microphone was mounted flush on a circular 1~m diameter rigid plate placed on the ground. A 9~cm diameter standard wind screen cut in half was used as a primary wind screen and a 75~cm diameter hemispherical secondary wind screen (Microtech Gefell type GFM920) was used to further reduce wind-induced noise. The rationale behind the choice of this microphone setup in the context of the preparation of a listening test was discussed elsewhere \cite{Dutilleux:2020aa}.
During the measurements, the wind speed was measured at 1.5~m height with a handheld anemometer (Windmate type WM-100). In addition, the wind farm managers provided the wind speed value at hub height over the hour corresponding to the measurements.

The recordings were made in September 2020 during daytime. Based on the cloud cover (2/8) on site, the conditions during the recordings belonged to the stability class 1 \cite{ISO:2017uo}. At both wind farms, the ground was a patchwork of gravel, rocks and low vegetation, such as moss or short grass a few centimeters high. No significant noise was produced by the sparse low vegetation and the recordings took place at sites far away from any other noise-generating infrastructure. Therefore, it can be considered that the noise recorded was clearly dominated by the wind turbines. 

\begin{table*}
  \begin{center}
    \begin{tabular}{p{5cm} c c}
      \hline
        Wind farm & Storheia & Bessakerfjellet \\
      \hline
        \textit{Description} & &\\
        Commissioning year & 2019 & 2008 \\
        Number of turbines & 80 & 25 \\
        Turbine manufacturer & Vestas & Enercon \\
        Turbine type & V117 & E70/2300\\
        Nominal electric power (MW) & 3.6 & 2.3 \\
        Hub height (m) & 87 & 64.5 \\
        Rotor diameter (m) & 117 & 71 \\
      \hline
        \textit{Recording campaigns} & & \\
        Wind speed during the recordings (m/s) at hub height & 9.7-10.7 & 7.0-8.0 \\
        Recordings selected for the listening test & \autoref{fig:micPosStorheia} & \autoref{fig:micPosBessaker} \\
      \hline
    \end{tabular}
  \end{center}
  \caption{Information about (1) the two wind farms where audio recordings were carried out and (2) the corresponding recording campaigns.}
  \label{tab:camp}
\end{table*}

\begin{figure}[h]
\figline{\fig{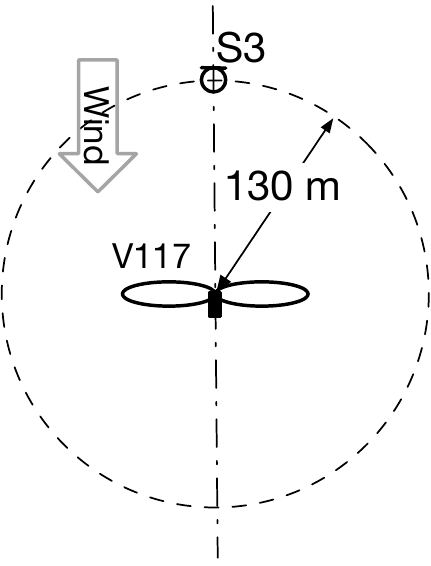}{4cm}{(a) Storheia}\label{fig:micPosStorheia}
\fig{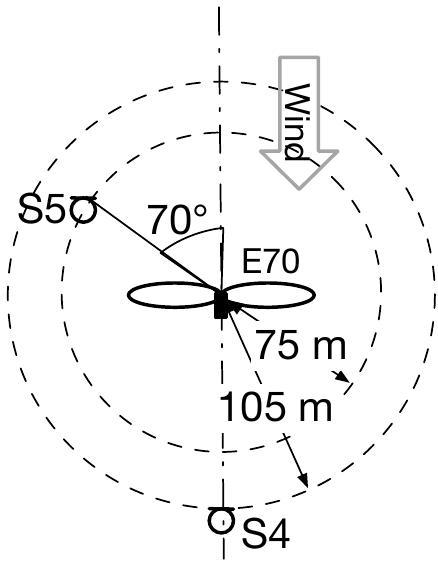}{4cm}{(b) Bessakerfjellet}\label{fig:micPosBessaker}}
\caption{Microphone positions for the recordings used in the listening tests.}
\label{fig:micPos}
\end{figure}

\subsection{Pre-processing wind turbine sounds}
\label{sec:preproc}

While wind turbine sounds were collected close to the turbines in order to maximize the signal-to-noise ratio, the aim was to prepare stimuli whose spectral content was closer to what can be experienced at a more realistic distance in a typical downwind scenario, where the wind farm is in direct line-of-sight of the receptor. The target distance was set to 1000~m. Therefore, the raw wind turbines sounds were pre-processed to account for spectral alterations during sound propagation. Considering that annoyance due to tonality is beyond the scope of our study,  tonalities were also removed from the recordings during the pre-processing. No correction for insertion loss due to the secondary wind screen was applied. 

The atmospheric attenuation at 1000~m was computed according to the ISO 9613-1 standard \cite{ISO:1993aa}. The corresponding filter was applied in the frequency domain. In order to take into account the influence of downward refraction and of ground effect, a frequency filter was created using an adaptation of the Bellhop ray-tracing software \cite{Porter:1987tx} to atmospheric sound propagation \cite{Hussain:2020aa}. For the needs of the simulation, the following additional assumptions were made: above a flat ground, the source height considered was that of the turbine recorded at Storheia (\autoref{tab:camp}) and the receiver was placed at 2~m height. A range-independent sound speed profile was calculated using Monin-Obukhov’s similarity theory for the W3S4 case \cite{ISO:2017uo}. A ground impedance model with variable porosity was assumed, with values for heath \cite{Attenborough:2011aa}. 

The occasional tonal components were first detected using both listening and a Python program \cite{Gle:2016aa} that implements a standard procedure for the detection of tonality \cite{ISO:2016ab}. The tonalities were then filtered out in the frequency domain using a procedure described elsewhere \cite{Dutilleux:2020aa}.

\subsection{Recording residual sounds}
The first residual sound $R1$ was a monophonic recording of vegetation, groves and high grass sounds made at Dragvoll (Trondheim, Norway) at 1~m height on a windy day of September 2020. Wind speed at 1~m height varied between 1.0 and 2.5~m/s. The recordings were made with an omnidirectional prepolarized 1-inch microphone capsule (AKG type CK2) mounted on a 48 V preamplifier (AKG type C460B) in a rigid-framed windshield (Rycote type Cyclone). Here, the recordings were not calibrated because a suitable calibrator mouthpiece was not available for the microphone chosen. Instead, the sound pressure level was monitored close to the recording point with a class 1 sound level meter (Norsonic type Nor150) protected by a standard 9 cm wind screen. The sound level meter was calibrated before and after the measurement session with a class 1 calibrator (Norsonic type Nor1256). The $L_{Aeq}$ for $R1$ was 44 dB.  
The audio signal was stored on a digital audio recorder (Sound Devices type 722) using a sampling frequency of 48~kHz and a resolution of 24 bits.

The second monophonic recording of residual sound $R2$ was collected at Berg (Trondheim, Norway) in a residential area on a windless day, in June 2019. The background soundscape was dominated by the city and the traffic on a  highway located several hundred meters away. The hardware used was a hand-held recorder (Sony type PCM D100) with built-in cardioid microphones, 44.1 kHz sampling frequency and 16-bit resolution. The recording was not calibrated during the collection of the second residual sound, but it was kept track of gain settings. A approximate calibration was later performed in NTNU's anechoic room (cut-off frequency 100 Hz) with a loudspeaker cabinet (Genelec type 1029A) radiating a 1 kHz tone. The recorder's microphones were located next to a that of a class 1 sound level meter (Norsonic type Nor150) at 1.75 m from the source. Based on this, the approximated $L_{Aeq}$ for $R2$ was 53 dB. 

\subsection{Selected residual and specific sound samples}

\begin{figure*}[h]
\ifthenelse{\boolean{color}}{
\figline{
\fig{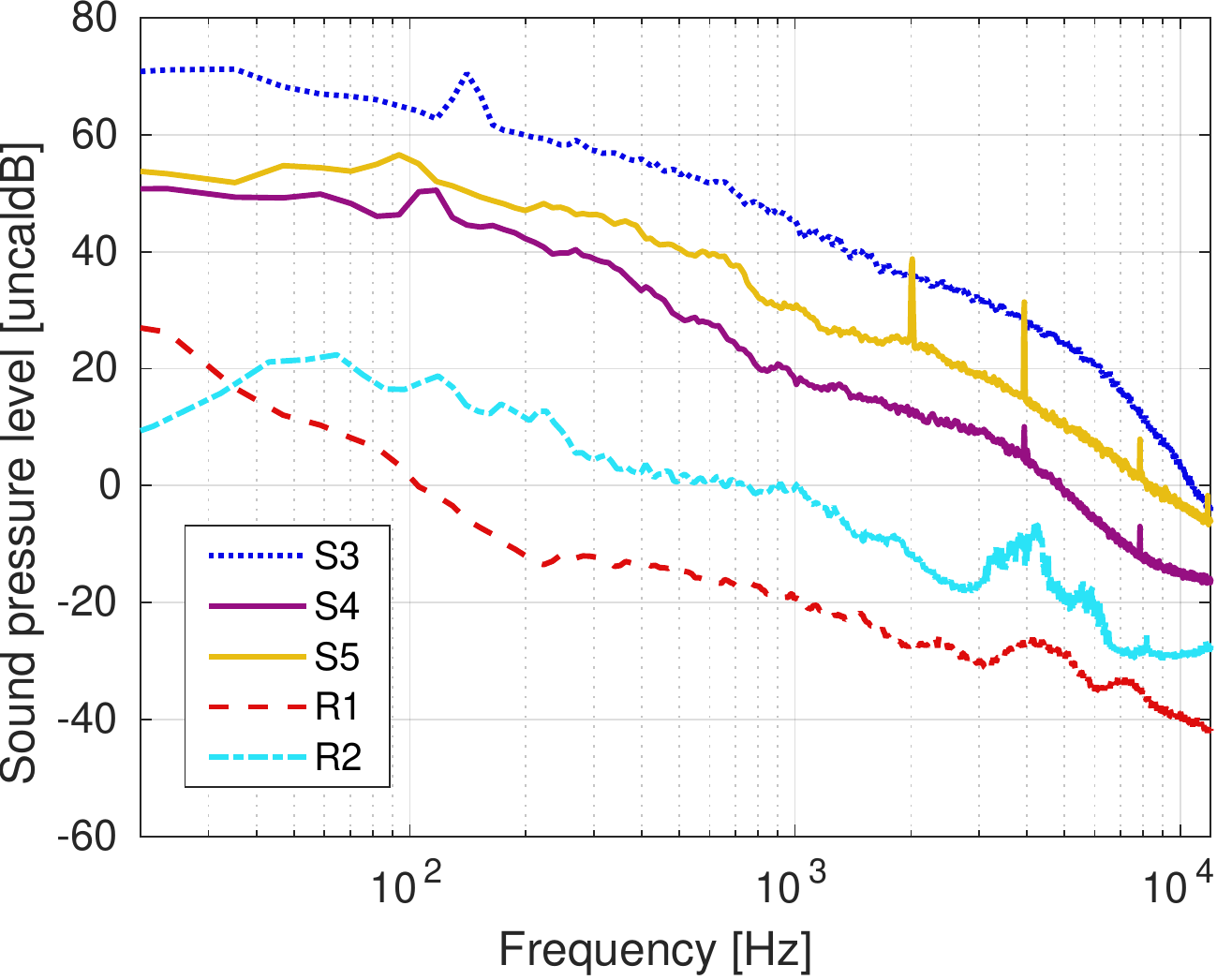}{5cm}{(a) Frequency spectra of specific and residual sounds}\label{fig:fft}
\fig{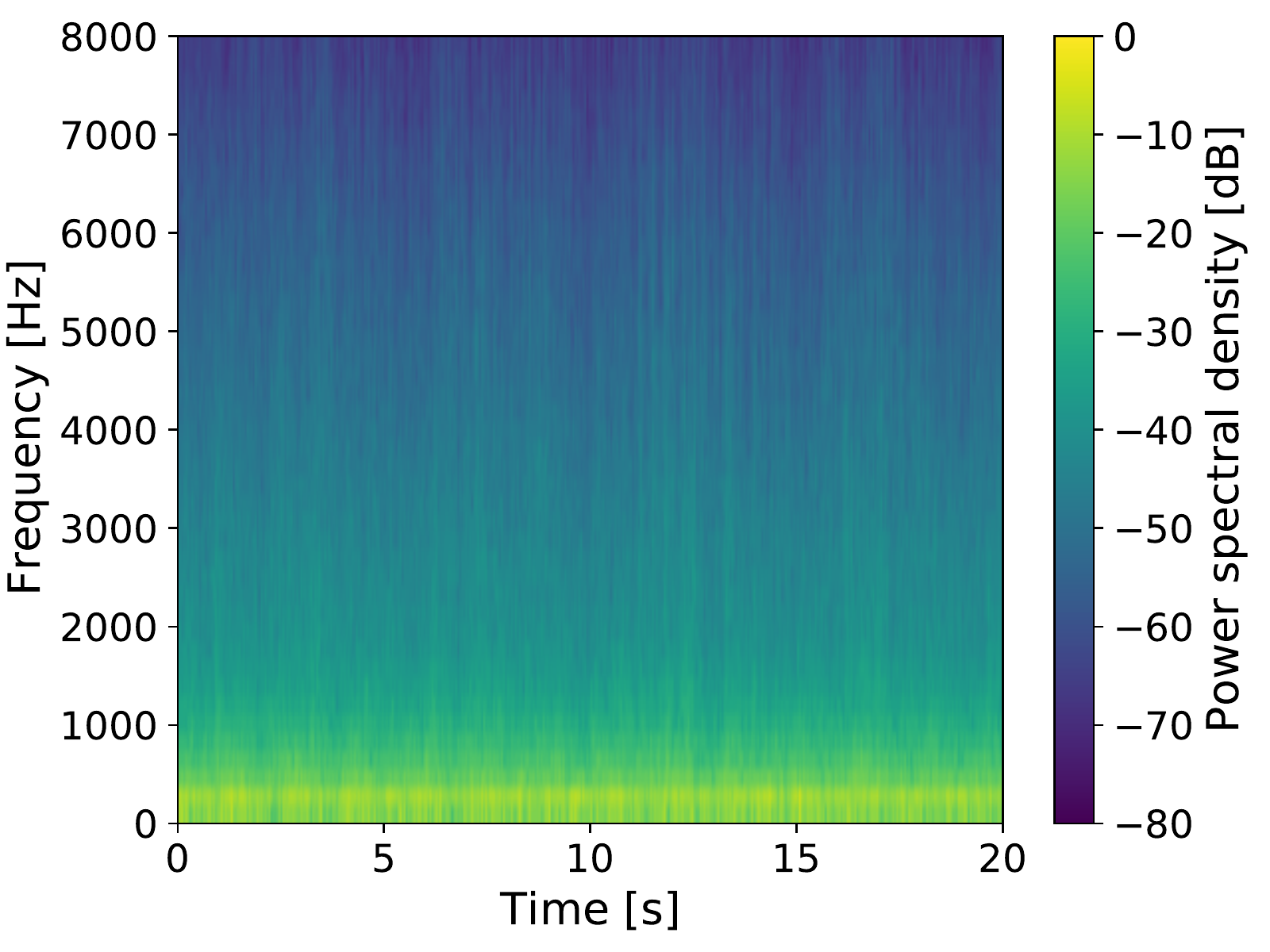}{6cm}{(b) Specific sound S3}\label{fig:S3}}
\figline{
\fig{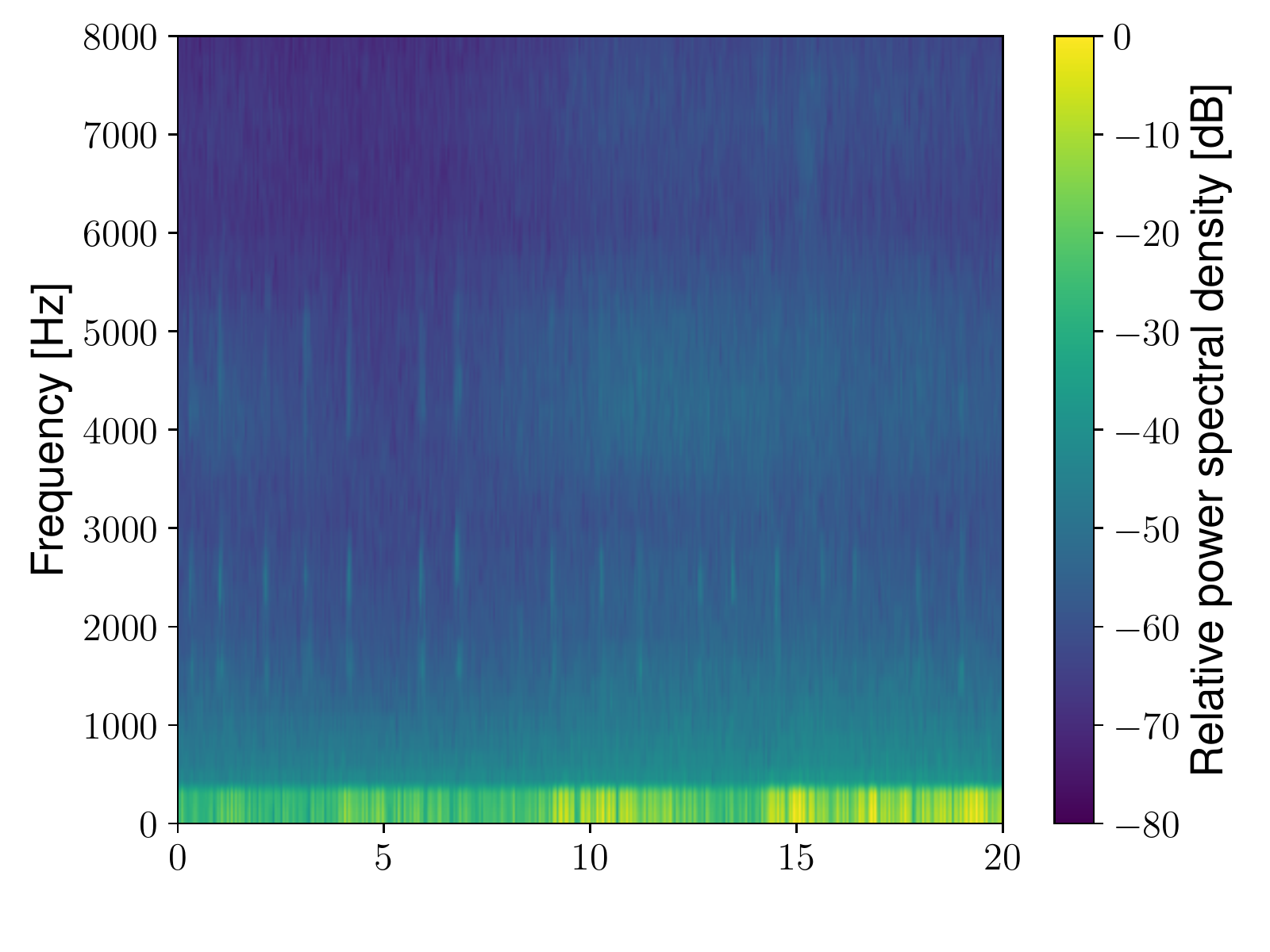}{6cm}{(c) Residual sound R1}\label{fig:R1}
\fig{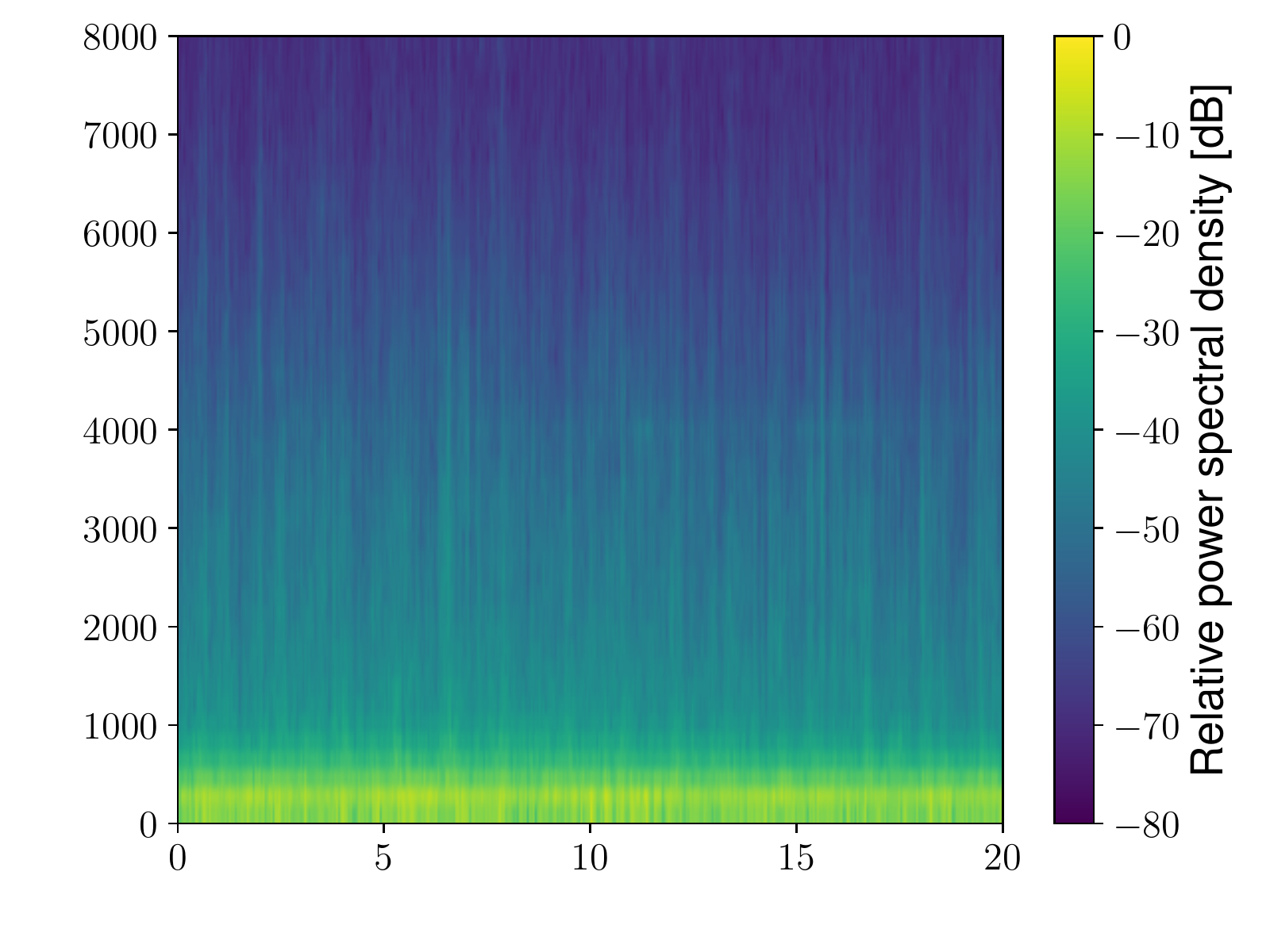}{6cm}{(d) Specific sound S4}\label{fig:S4}}
\figline{
\fig{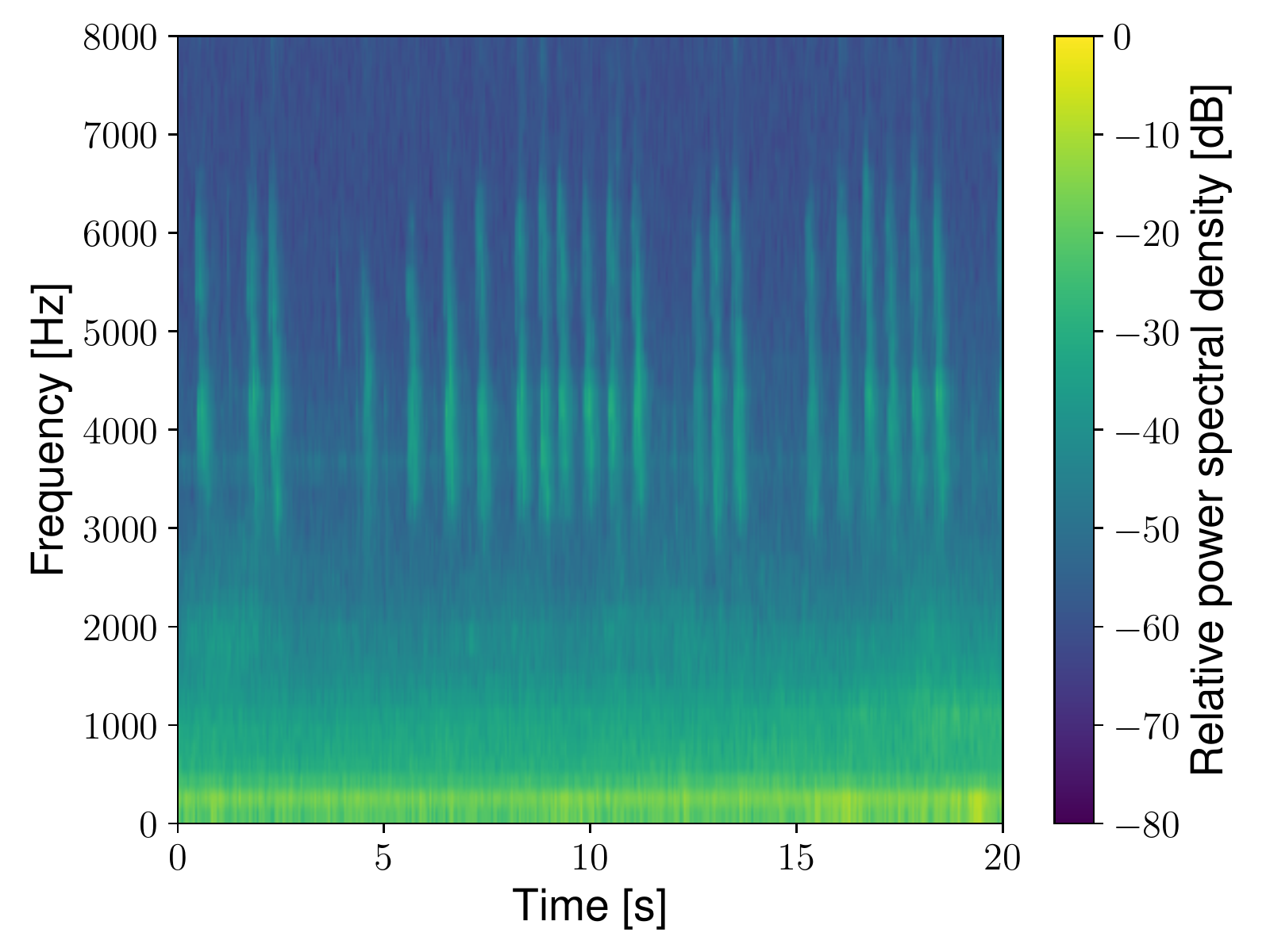}{6cm}{(e) Residual sound R2}\label{fig:R2}
\fig{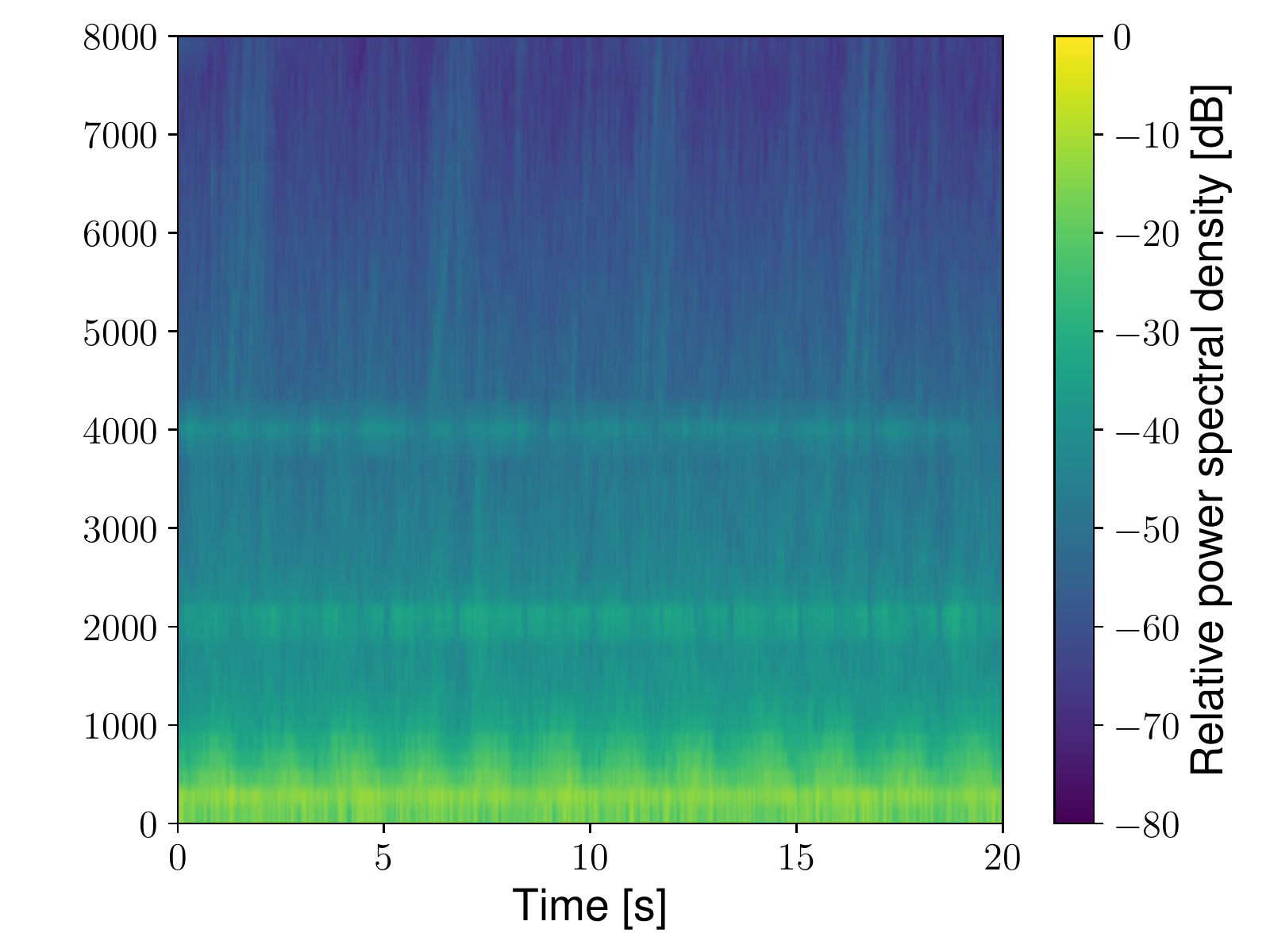}{6cm}{(f) Specific sound S5}\label{fig:S5}}
}
{
\figline{
\fig{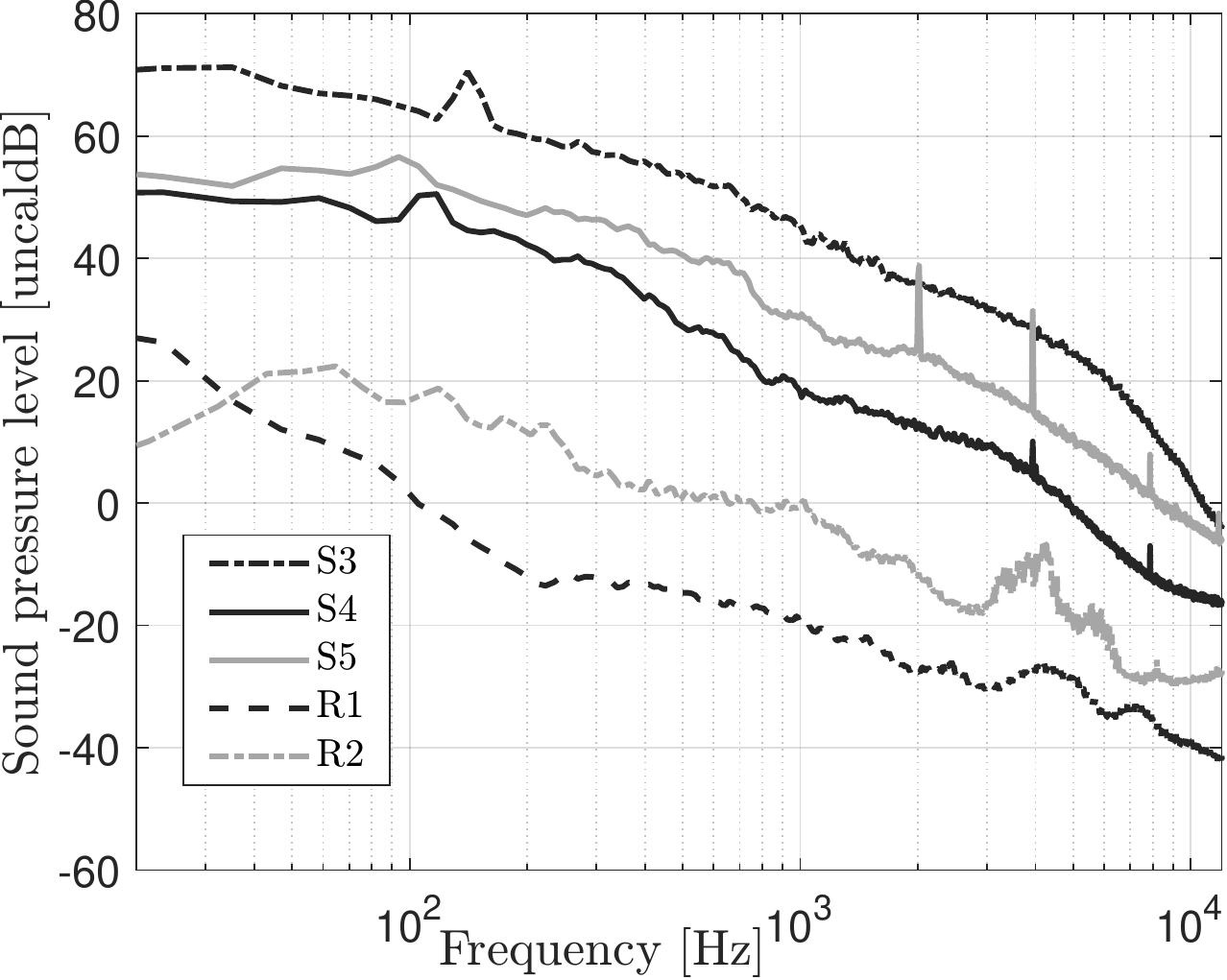}{5cm}{(a) Frequency spectra of specific and residual sounds}\label{fig:fft}
\fig{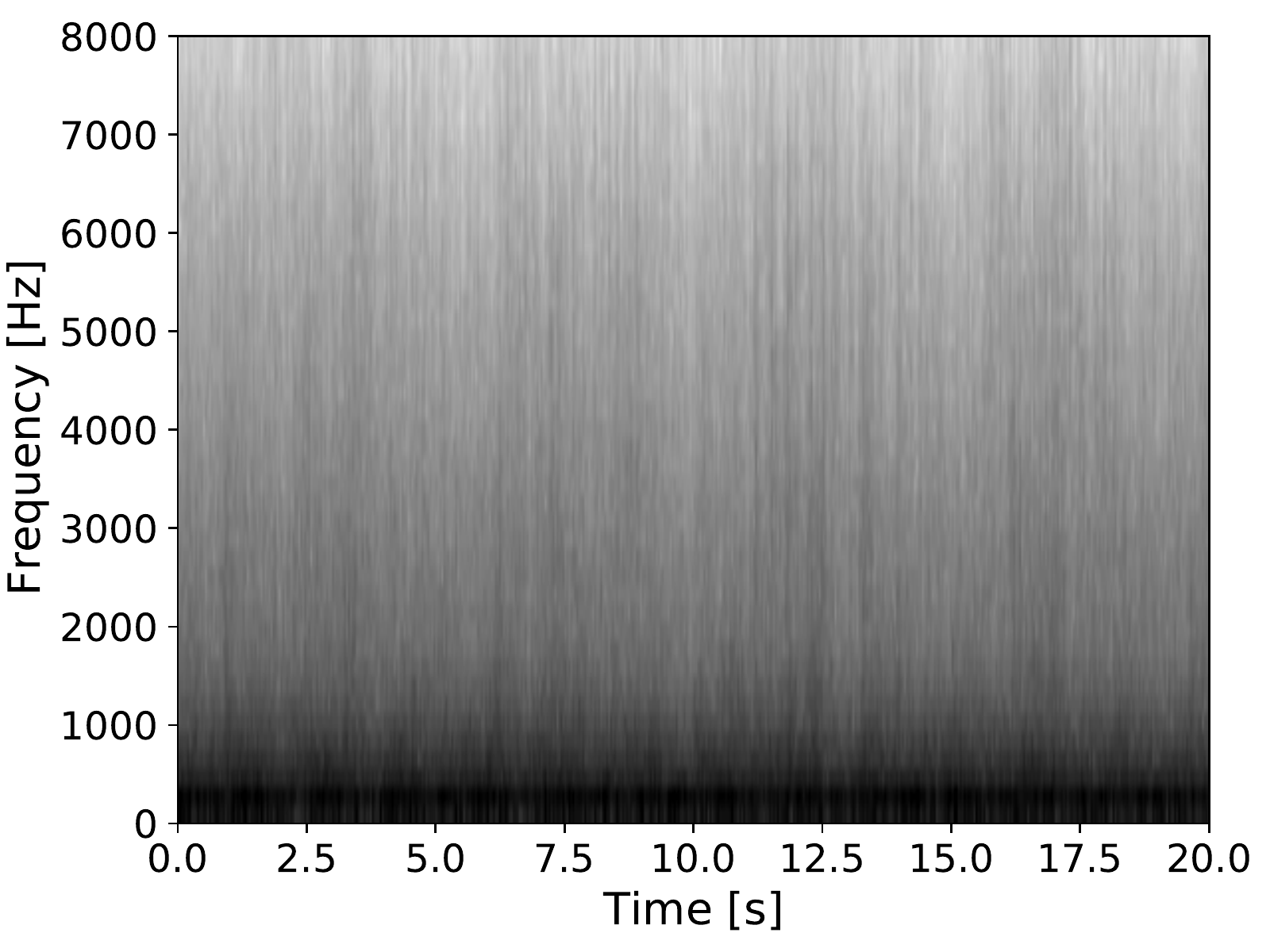}{6cm}{(b) Specific sound S3}\label{fig:S3}}
\figline{
\fig{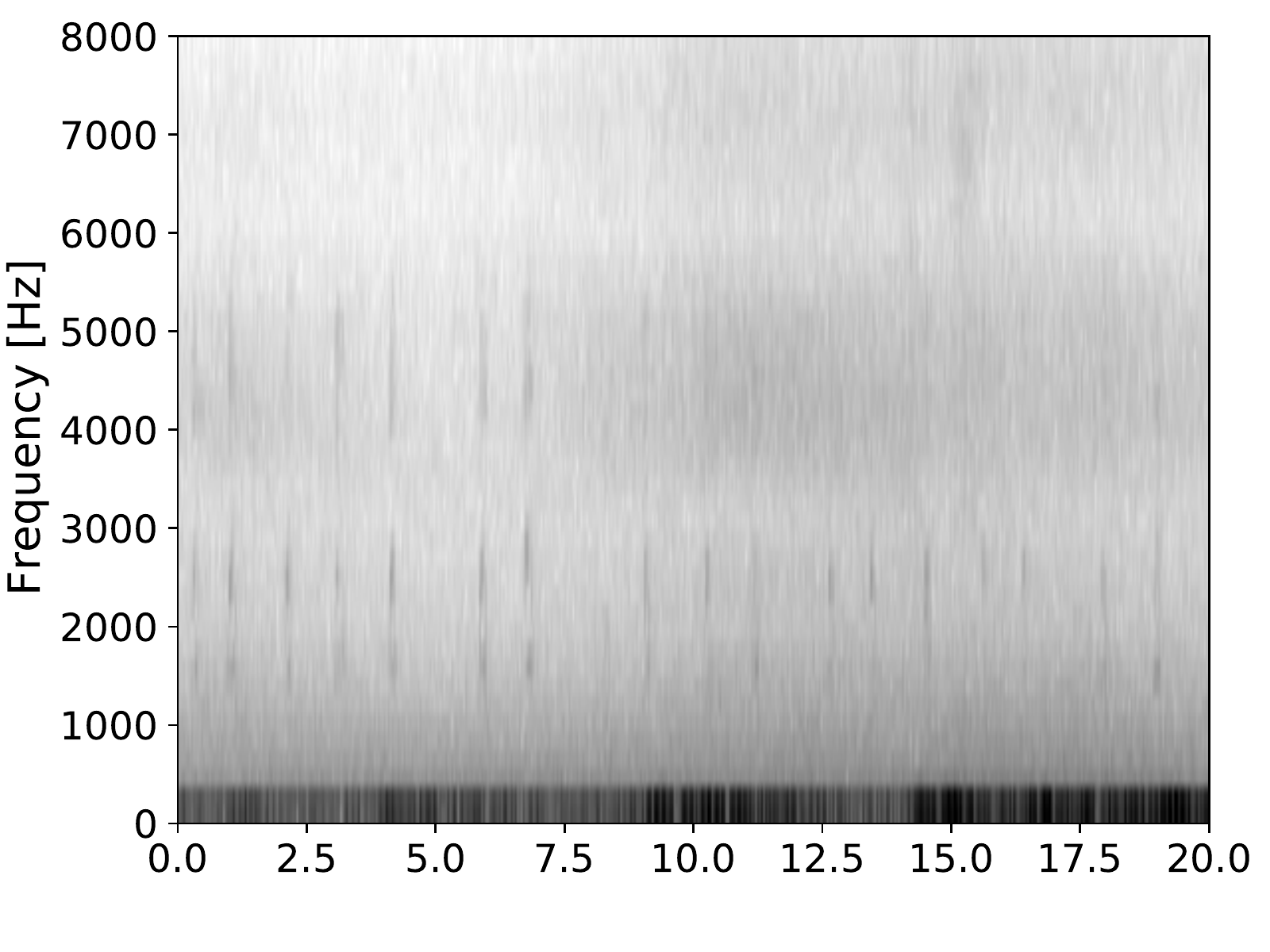}{6cm}{(c) Residual sound R1}\label{fig:R1}
\fig{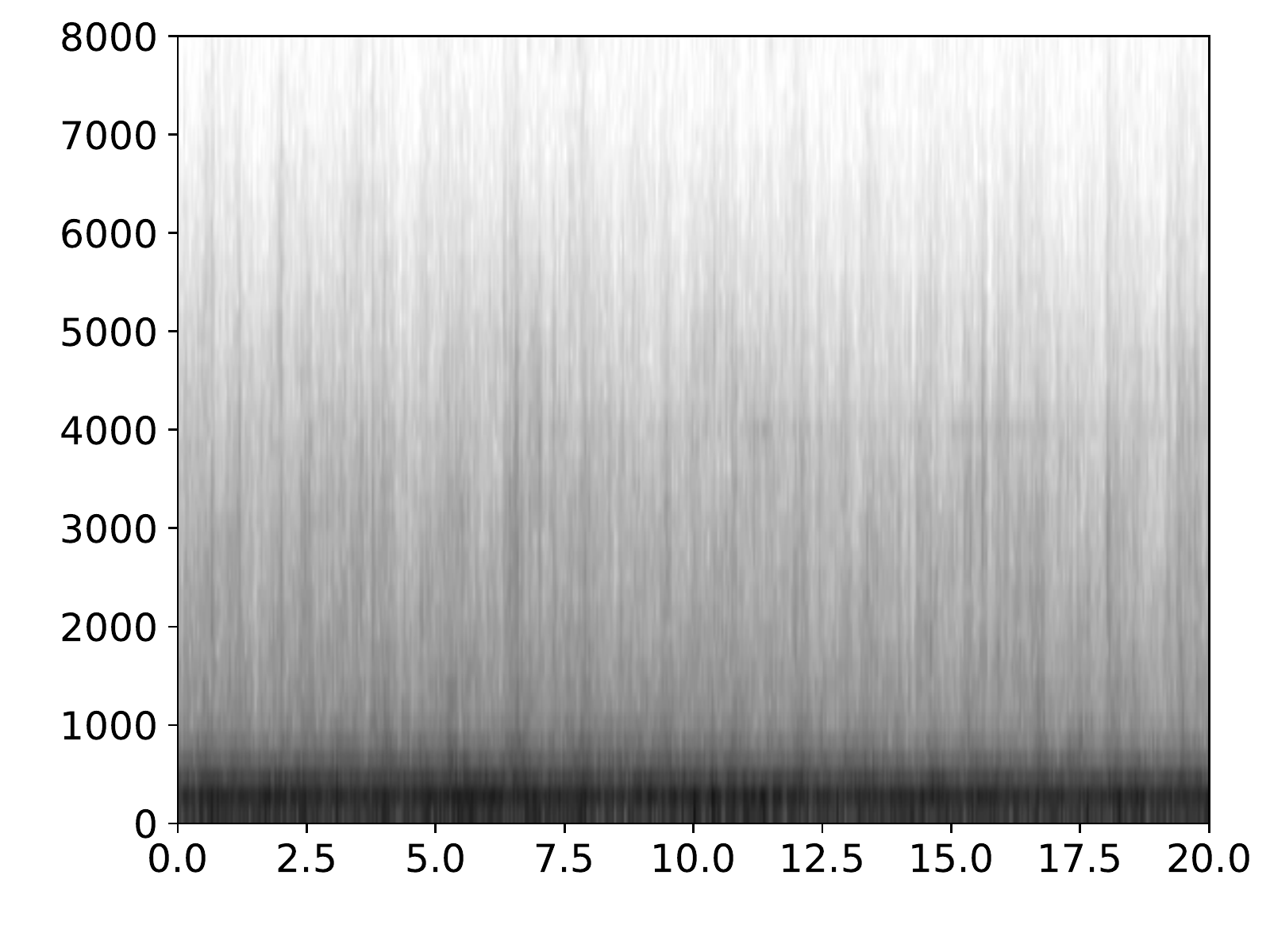}{6cm}{(d) Specific sound S4}\label{fig:S4}}
\figline{
\fig{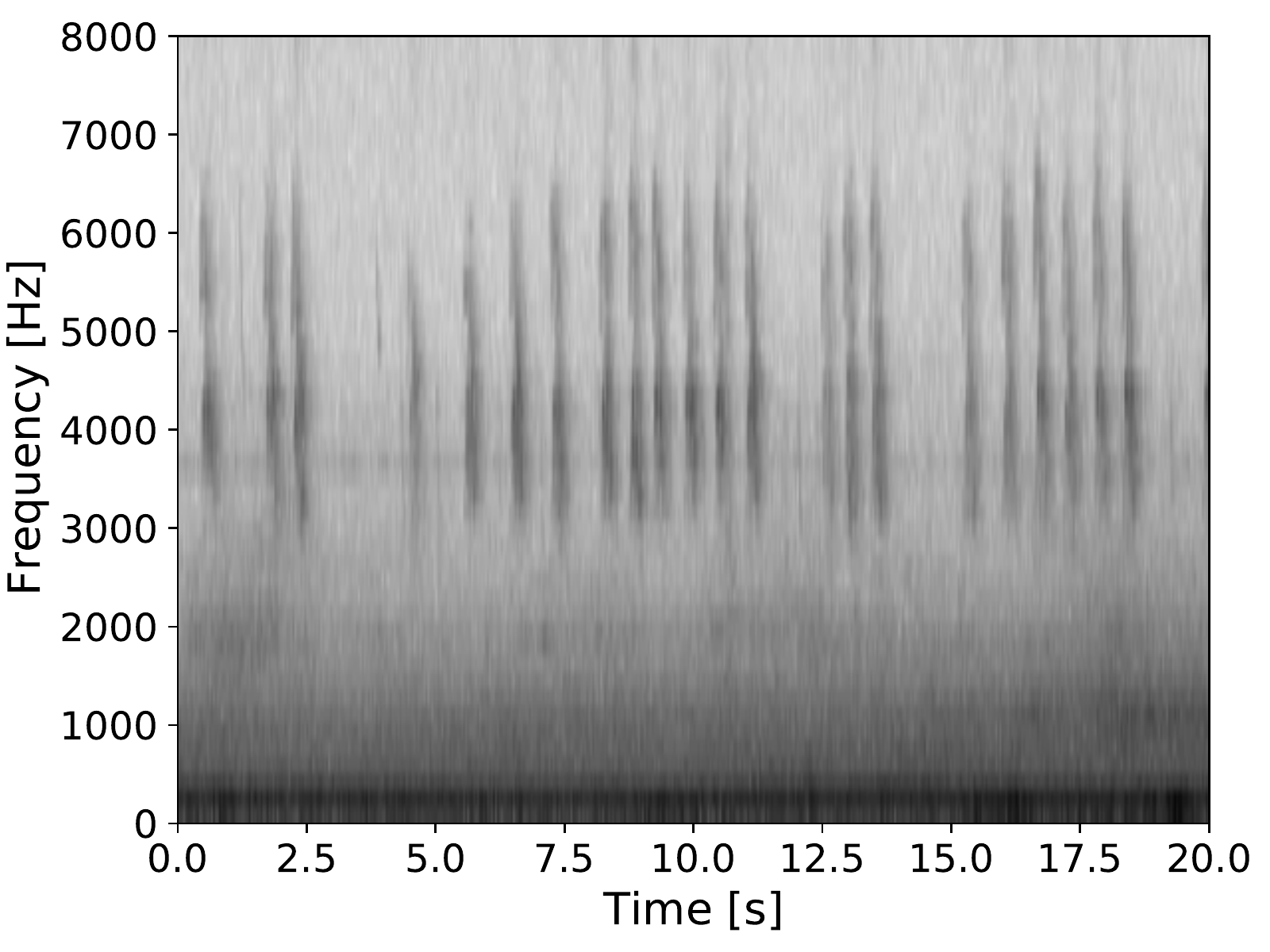}{6cm}{(e) Residual sound R2}\label{fig:R2}
\fig{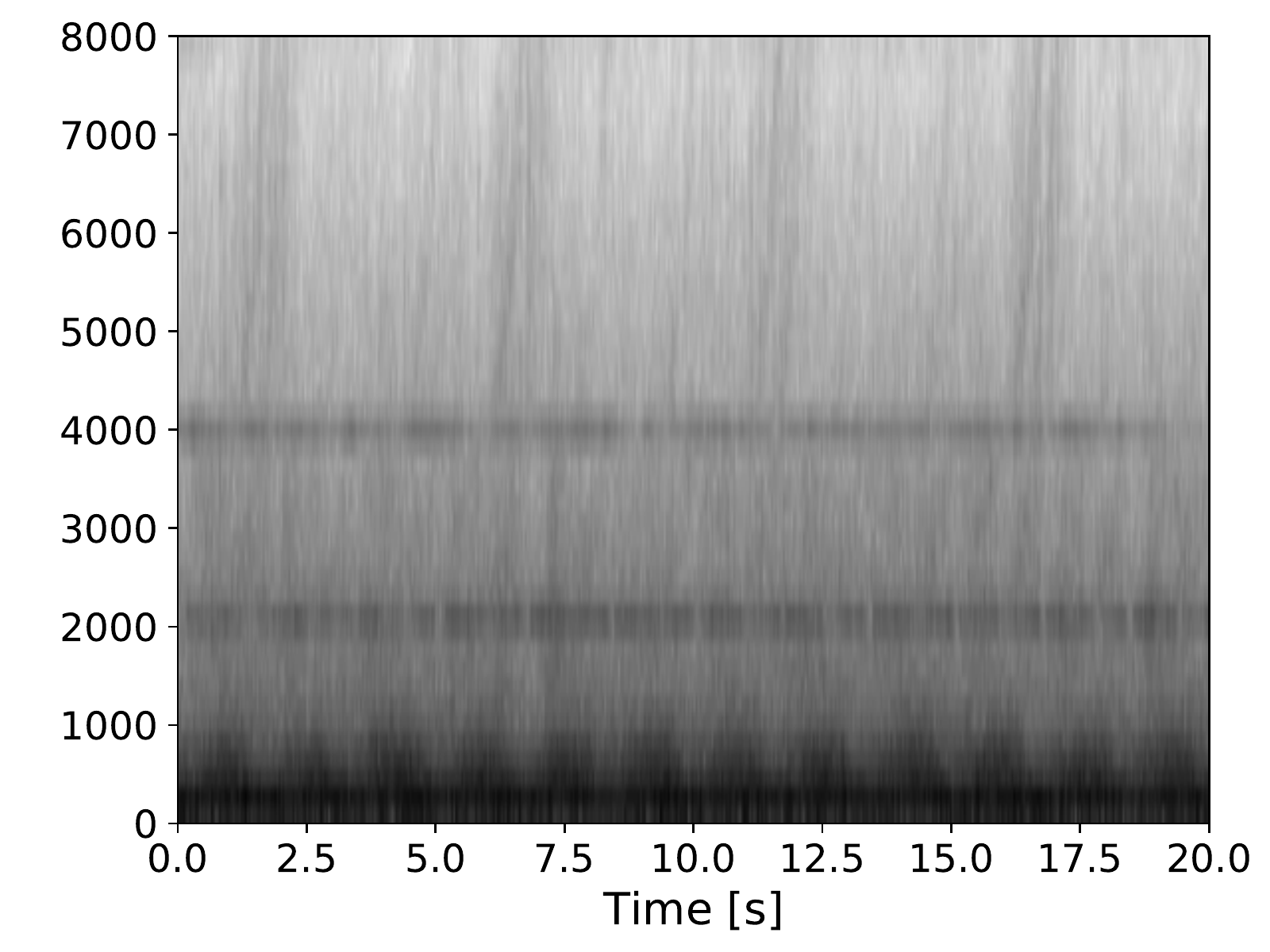}{6cm}{(f) Specific sound S5}\label{fig:S5}}
}
\caption{ Overview of the signals selected to build the stimuli for the listening tests. For the specific sounds, the figures are based on the raw signal, \textit{i.e.}  before correcting for atmospheric attenuation, the ground effect and removing occasional tonal components. In (a), the magnitudes of the frequency spectra are arbitrary. 
The light tonality present in $S4$ ((a) and (d)) can be seen around 4 kHz. (a) and (f) for $S5$ show tonal components around 2 and 4 kHz and the turbine's swish. In the residual sounds $R1$ (c) and $R2$ (e), the transients are bird vocalizations. The tonality below 4 kHz attributable to a heat pump can be seen in (a) and (e) for $R2$ .}
\label{fig:spectres}
\end{figure*}

The four specific sounds $S3$, $S4$, $S5$ and $S6$ were selected from the recordings of wind turbine sounds to generate the stimuli: $S3$ was recorded in upwind position at the Storheia wind farm, $S4$ corresponds to a downwind position at Bessakerfjellet and $S5$ was recorded at another wind turbine at Bessakerfjellet from a 70° angle with the wind direction (\autoref{tab:camp}). The specific sound $S6$ originates from the same recording as $S5$ but was post-processed slightly differently. For this sound clip, the propagation was simulated without ground effect in order to assess the influence of this phenomenon on short-term annoyance. An overview of the characteristics of the three raw wind turbine sounds \emph{before} the corrections for tonality and sound propagation is given in figure~\ref{fig:spectres}.

All the raw wind turbine sounds contained tonalities at frequencies ranging from 130~Hz to 8~kHz. The procedure used was effective at removing them. In addition, a clear fluctuation or "swish" can be heard in $S5$ (\autoref{fig:S5}) and $S6$ , whereas $S3$ (\autoref{fig:S3}) and $S4$ (\autoref{fig:S4}) can be described as more steady sounds.

The residual sound $R1$, which corresponds to the sound of wind in the vegetation, had a very broadband spectrum and no tonalities. It mainly contained broadband sound from the wind in the leaves and tall grass, combined with sporadic calls from an Eurasian magpie (\textit{Pica pica}). In $R2$, traffic noise could be heard in the distance as well as House sparrows (\textit{Passer domesticus}) calling nearby. Two tonalities around 4.0 and 10.2 kHz were slightly audible and generated by a heat pump. The PSD of the signal and a spectrogram are presented in figure~\ref{fig:spectres}. While $R1$ and $R2$ were collected in Norway, both residual sounds could have been recorded in many other places across Europe. In particular, the two bird species that can be heard are common species with a wide distribution area. 

The specific and residual sound selected are subjectively free from any wind-induced noise on the microphone. 



\subsection{Combining residual and specific sounds}
As explained in section~\ref{sec:preproc}, the wind turbine sounds were cleaned from any occasional tonal component and their frequency spectrum was extrapolated to a distance of 1000~m. The specific sound was then combined with the residual sound to create a range of stimuli with a specified sound emergence $e$ (Eq. \ref{eq:e}) and a specified total sound pressure level $L_{\text{tot}}$. The metric used for evaluating both $L_{\text{tot}}$, $L_{\text{spec}}$ and $L_{\text{res}}$  is the $L_{Aeq}$, computed over the duration of the sound stimulus. 

In the light of typical values for $L_{\text{tot}}$ \cite{Hongisto:2017up, Michaud:2016va} and limit values for $e$ at facades exposed to wind turbine sounds, three $L_{\text{tot}}$ values were chosen at $30, 40$ and $50$~dB(A) and four \text{SRR} (Eq. \ref{eq:SRR}) values were fixed at -10, -5, 0 and 5~dB. The latter values correspond to $e$ at 0.4, 1.2, 3.0 and 6.2~dB to the nearest tenth of a dB. SRR was deemed more convenient than $e$ when it comes to preparing the stimuli because the range of variation of $e$ is limited to 3 dB for negative SRR. Each listening test was based on only one specific residual sound. $T1$ was based on $R1$, and $T2$ on $R2$. This means that for both listening tests, 48 different combinations of specific sound, $L_{\text{tot}}$ and $e$ were used.  

\subsection{Test facility}
The room selected for the listening test is the receiving room of a decommissioned test facility for the measurement of the sound insulation of building materials. It is a shoebox-shaped room of dimensions $5.87\times4.88\times4$~m. The room is vibration-isolated from the rest of the building where it is located. The only openings are double heavy steel doors that were closed during the listening tests. The room has no window and is visually neutral. During the experiment the room was occupied with various lab equipment. 

The A-weighted equivalent sound pressure level in the room was measured in May 2021 with a calibrated class I sound level meter (Norsonic type Nor150) coupled with a 1-inch low self noise condenser microphone (Bruel \& Kjær type 4144) and a 200 V pre-amplifier (Norsonic type Nor1201) during the same time slots than the ones used for the listening test. The $L_{Aeq,1s}$ showed only minor fluctuations. With an average A-weighted background sound pressure level at 17.5 dB the room was deemed suitable to play back stimuli at 30 dB through headphones. 

The listening test relied on software developed in a previous research project and was described elsewhere \cite{Dutilleux:2020aa}. The user interface does not perform any processing on the acoustic stimuli that are prepared beforehand and stored in mono PCM 48 kHz 24 bits format.

The calibrated sound reproduction chain consisted of (1) a laptop computer (Dell type Vostro), (2) an external sound card (Roland Studio Capture type UA1610) and a pair of open circumaural electrodynamic headphones (Beyerdynamic type DT990 Pro). For the sake of the listening test, the frequency response of the headphones was measured with a dummy head (Neumann type KU100) and an inverse filter was applied to equalize the frequency response from 100 to 6000 Hz. In that range, the corrected frequency response of the headphone followed standard guidelines for audiometry \cite{IEC:2017va}.

\subsection{Listening test protocol}
The software user interface guides the subject through the listening test. Starting with the collection of personal data, it proceeds with the presentation and the rating of the different stimuli and the storage of the results for further processing. The software controls the reproduction of the different stimuli in a random order which is varied from one subject to the next. The user interface also controls the audio hardware while collecting the subject's answers regarding short-term annoyance and the audibility of wind turbine sounds for each stimulus. Each subject listens to the stimuli one by one. The task is introduced by ``\textsl{Imagine that you live nearby a wind farm and that sometimes, when you are outside or have a window open, you hear this sound. How disturbing would you experience the sound to be on a scale from 0 - 10. Mark your unpleasantness}.'' The subject can decide when to play each sound stimulus but is not allowed to replay it or to go backward. Right after the sound stimulus was played back, the subject rates the short-term annoyance using the standard ISO 15666 \cite{ISO:2003aa} 11-point scale that is displayed by the graphical user interface. Then, the subject answers whether or not he/she could hear a wind turbine in the stimuli just played back, before proceeding to the next sound stimulus.

As for the collection of personal data, the protocol was approved by the Norwegian Center for Research Data (NSD - \textit{Norsk Senter for Forskningsdata} - application number 813181). 

\subsection{Test subjects}

The subjects were recruited mostly among students from the Norwegian University of Science and Technology (NTNU, Trondheim, Norway) via a mailing list and social media announcements. The subjects were not asked to take an audiometric test but to self-report any hearing impairment.

For each listening test, 30 subjects with self-declared normal hearing were recruited. The two groups of 30 subjects were separated sets. There were 11 female subjects for 19 male ones in listening test 1 ($T1$) and 10 female subjects for 20 males in the second listening test ($T2$). The age of the subjects ranged from 20 to 35 years old with a median age of 23.5 years in $T1$ and quite similarly the subjects ranged from 20 years to 41 years with a median age of 25 years in $T2$. 14 nationalities (among them 78~\% from Europe) and 13 nationalities (with 57~\% from Europe) were represented in the panel of subjects in, respectively, $T1$ and $T2$. 10 and 7 participants reported a prior experience of wind turbine sounds in each listening test. $T1$ was carried in April 2021 and $T2$ in May 2021. 

\subsection{Statistical analyses}
During the post-processing, we carried out an ANOVA test to examine the effects of sound emergence, the sound pressure level, the residual sound and the different specific sounds on short-term annoyance. The Greenhouse–Geisser correction was used to adjust for lack of sphericity. In addition, we carried out linear regressions with short-term annoyance as the dependent variable and one or two noise metrics as the independent variables. In the statistical tests, a $p$-value lower than 0.05 was considered as statistically significant.  A complementary t-test was also realized to characterize the difference between the stimuli with a Bonferroni correction.

The statistical analysis was carried out in parallel by each co-author, either in R-4.2.0 (using the afex package release 1.1.1) or in Julia (version 1.5 with the packages ANOVA 0.1.0 and GLM 1.1.1). The computation of loudness-related metrics was carried out with the software implementation available from \cite{ISO:2017ww}.

\section{Results}
\label{sec:results}

\subsection{Short-term annoyance}

The ANOVA test showed that the influence of the residual sound is not statistically significant for short-term annoyance ($p=.48$), see \autoref{tab:ANOVA}. Thus, the data of both test $T1$ and $T2$ is treated simultaneously in the ANOVA allowing to increase the power of the analysis.

\begin{figure*}[h]
\figline{\fig{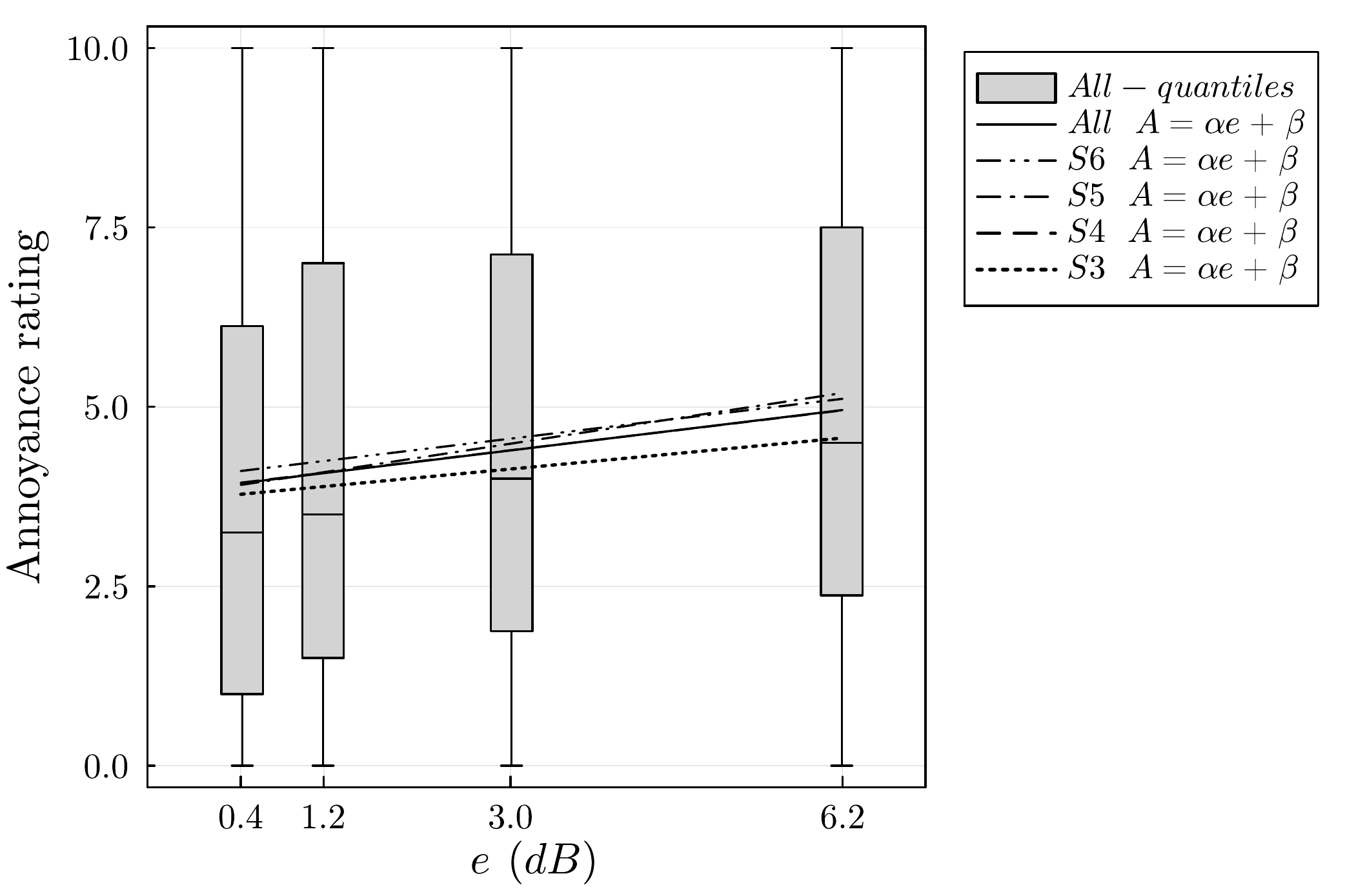}{9.5cm}{(a) T1}\label{fig:SNR 1}
\fig{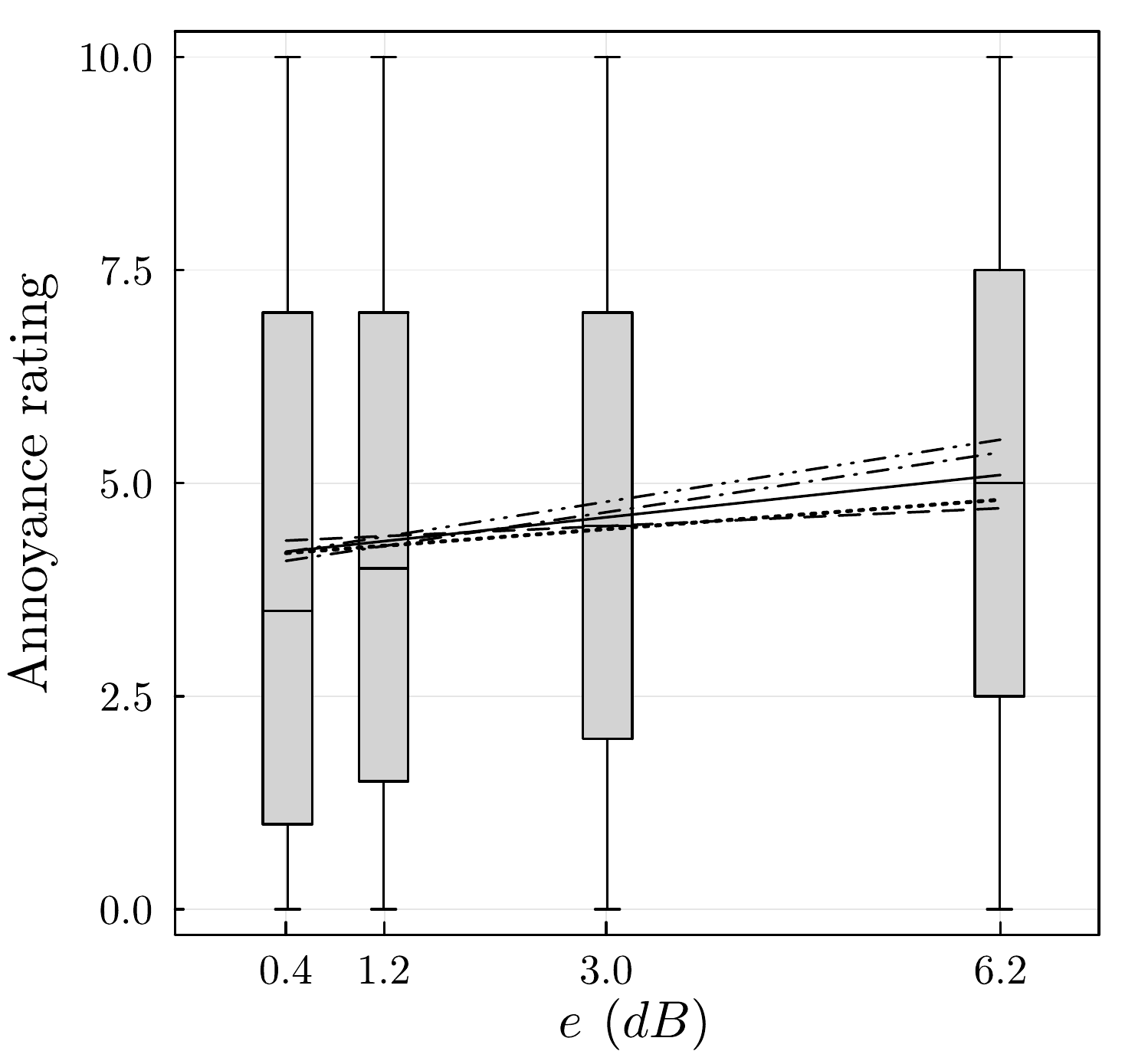}{7cm}{(b) T2}\label{fig:SNR 2}}
\caption{Short-term annoyance rating as a function of sound emergence.}
\label{fig:AnnVsEm1}
\end{figure*}

 Figure~\ref{fig:AnnVsEm1} presents box and whisker plots of short-term annoyance for the four values of $e$ (see \autoref{eq:e}) considered in this study. In both listening tests, both the progression of the median and the linear regressions show that short-term annoyance is a weakly increasing function of $e$. The associated positive slopes correspond to one point in the scale of annoyance per 5 dB increase in $e$ and the $\beta^{\ast}$ equal respectively 0.13 and 0.14. There is a large scatter in the annoyance ratings, with an interquartile range always larger than 4.8 units on the annoyance scale. In addition, the percentage of variance explained is low ($R^2=0.02$). All the details of the regressions are summarized in \autoref{tab:my-table}. 
 The ANOVA test confirmed a statistically significant effect of $e$ on short-term annoyance ($p<.001$, $F(1.6, 90.2)=38.2$), see \autoref{tab:ANOVA}.

\begin{table*}[h]
\begin{tabular}{rccccc}
\textbf{}             & $(\beta_i)$    & \textbf{$\alpha$}  & $(\beta_i^{\ast})$   & \textbf{RMSE}               & \textbf{$R^2$}               \\ \hline
$\beta \cdot L_{\text{tot}}+\alpha$    & 0.29 | 0.30     & -7.4 | -7.6                  & 0.79 | 0.80                                                                      & 1.89 | 1.85                 & 0.62 | 0.64                  \\ \hline
$\beta \cdot e+\alpha$                 & 0.18 | 0.16     & 3.9 | 4.1                    & 0.13 | 0.14                                                                      & 3.03 | 3.08                 & 0.02 | 0.01                  \\
$\beta\cdot e_{\text{UAC}}+\alpha$     & 0.25 | 0.25     & 3.7 | 3.7                    & 0.19 | 0.19                                                                               & 3.00| 3.00                  & 0.04 | 0.04                  \\
$\beta\cdot e_{\text{SP}}+\alpha$      & 0.12 | 0.06     & 3.8 | 4.3                    & 0.12 | 0.10                                                                                & 3.04 | 3.09                 & 0.01 | 0.00                  \\ \hline
$\beta\cdot \text{SRR} +\alpha$        & 0.07 | 0.06     & 4.5 | 4.7                    & 0.13 | 0.14                                                                      & 3.03| 3.08                  & 0.02 | 0.01                  \\ \hline
$\beta\cdot L_N+\alpha$                & 0.23 | 0.24     & -9.5 | -9.6                  & 0.78 | 0.79                                                                               & 1.91 | 1.89                 & 0.61 | 0.63                  \\
$\beta\cdot L_{N_{\text{max}}}+\alpha$ & 0.25 | 0.27     & -11.5 | -12.6                & 0.78 | 0.79                                                                                & 1.91 | 1.91                 & 0.61 | 0.62                  \\
$\beta\cdot L_{N_5} + \alpha $         & 0.25 | 0.25     & -10.8 | -11.12               & 0.78 | 0.79                                                                               & 1.91 | 1.89                 & 0.61 | 0.63                  \\ \hline
$\beta_L\cdot L_{\text{tot}}+$        & (0.29, | (0.30, & \multirow{2}{*}{-7.9 | -8.0} & \multirow{2}{*}{\begin{tabular}[c]{@{}c@{}}(0.79, | (0.80, \\ 0.13) | 0.14)\end{tabular}} & \multirow{2}{*}{1.8 | 1.82} & \multirow{2}{*}{0.63 | 0.65} \\
$\beta_e\cdot e+\alpha$                & 0.18) | 0.16)   &                              &                                                                              &                             &          
\end{tabular}
\caption{Results from the linear regressions. In each cell the value on the left (resp. right) is for $T1$ (resp. $T2$). This applies also to the last row that corresponds to a model with two independent variables ($\beta_L\cdot L_{\text{tot}}+\beta_e\cdot e+\alpha$): the value above is for $\beta_e$ and the value below for $\beta_L$.}
\label{tab:my-table}
\end{table*}

\begin{table*}[h]
\begin{tabular}{lccccc}
\hline
\textbf{Variable}                                            & \textbf{d.f} & \textbf{MSE} & \textbf{F} & \textbf{$\eta^2$} & \textbf{p-value} \\ \hline
Residual sound                                   & 1, 58        & 64           & 0.5        & \textless{}.01 & .48              \\
$L_{\text{tot}} $                & 1.5, 87.3    & 14           & 811.2      & .65            & \textless{}.001  \\
$e $                                                 & 1.6, 90.2    & 7            & 38.2       & .04            & \textless{}.001  \\
Specific sound                                      & 2.1, 124.3   & 3            & 7.2        & \textless{}.01 & \textless{}.001  \\
Residual sound :~$ L_{\text{tot}}$ & 1.5, 87.3    & 14           & 0.2        & \textless{}.01 & .77              \\
$L_{\text{tot}}$ : $e $               & 5.2, 302.6   & 2            & 4.5        & \textless{}.01 & \textless{}.001  \\
Residual sound : Specific sound                   & 2.1, 123.3   & 3            & 0.6        & \textless{}.01 & 0.55             \\
Residual sound : $ e $                               & 1.6, 90.2    & 7            & 0.2        & \textless{}.01 & .80              \\
Specific sound : $e $                                 & 7.8, 451.5   & 2            & 2.6        & \textless{}.01 & .01              \\
Specific sound : $L_{\text{tot}}$ & 5.1, 296.9   & 2            & 0.7        & \textless{}.01 & .63              \\ \hline
\end{tabular}
\caption{Results from the ANOVA test, $\eta^2$ is the generalized effect size.}
\label{tab:ANOVA}
\end{table*}

\begin{figure*}[h]
\figline{\fig{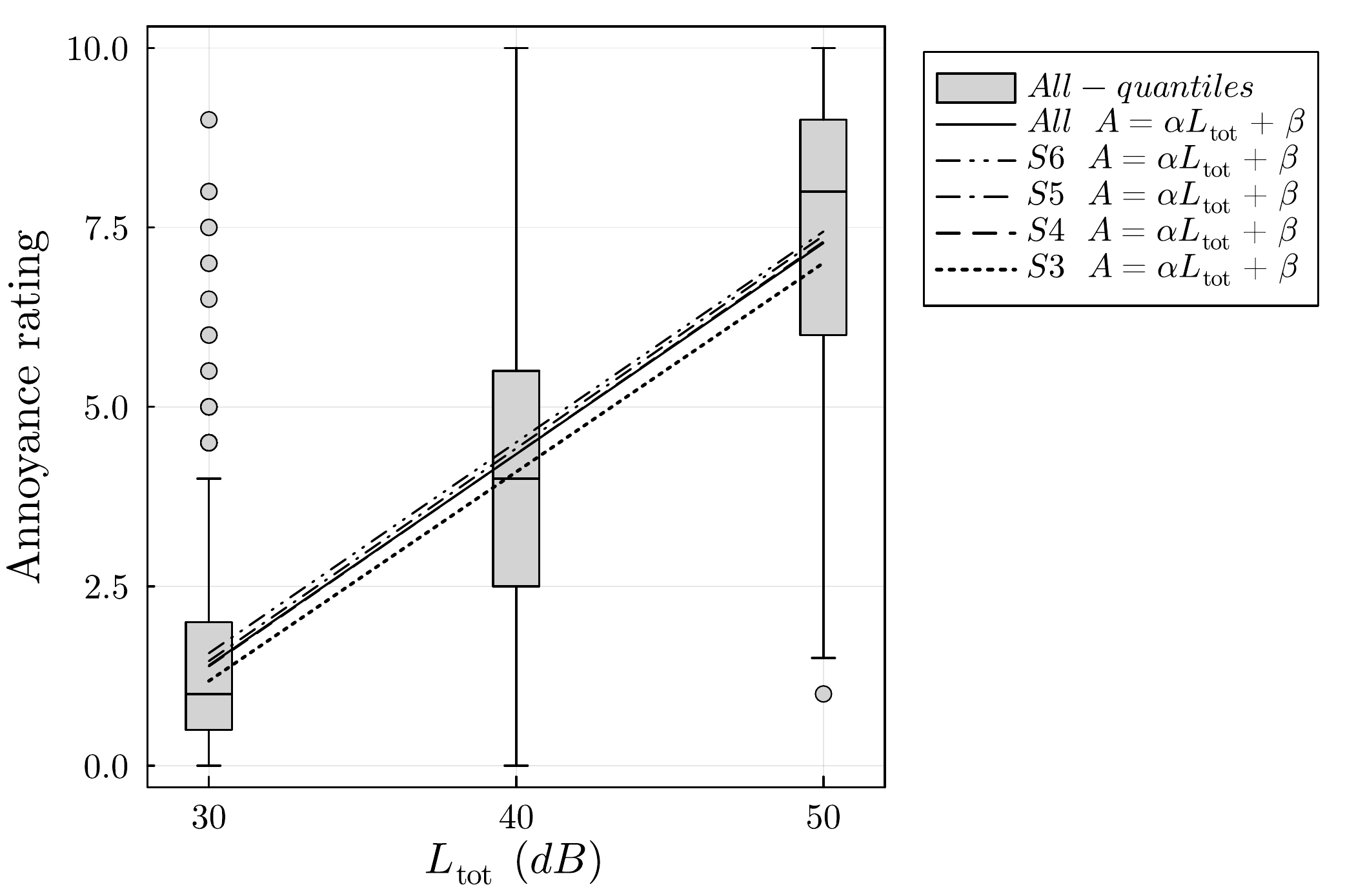}{9.5cm}{(a) T1}\label{fig:SPL 1}
\fig{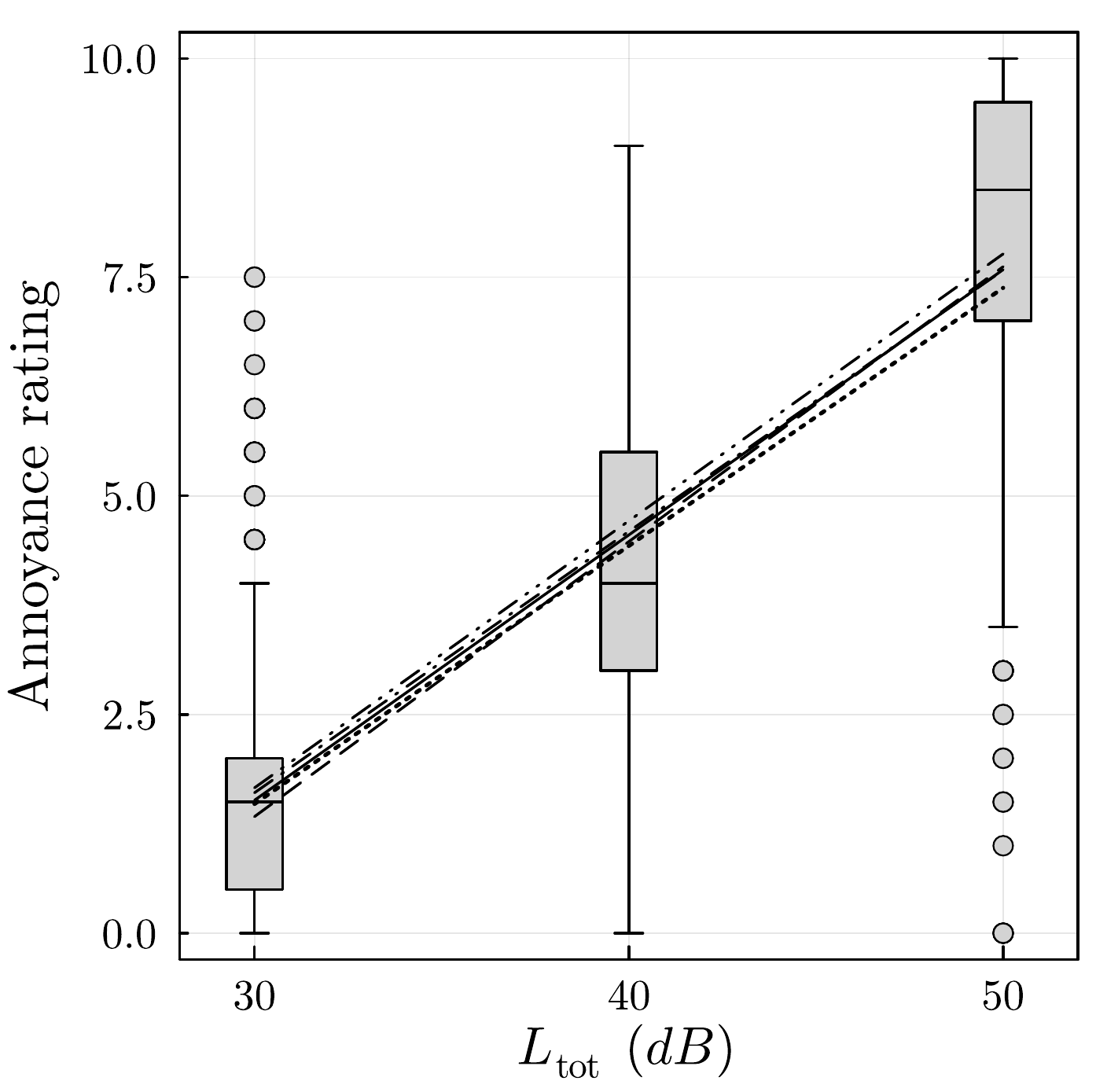}{7cm}{(b) T2}\label{fig:SPL 2}}
\caption{Short-term annoyance rating as a function of the total sound pressure level.}
\label{fig:AnnVsLp1}
\end{figure*}

The same analysis was carried out for short-term annoyance and $L_{\text{tot}}$. The ANOVA returned a significant effect of 
$L_{\text{tot}}$  ($F(1.5,87.3)=811.2$, $p < .001$). The corresponding box and whisker plots for the three values of $L_{\text{tot}}$ considered in this study are presented in \autoref{fig:AnnVsLp1}. The results of the linear regression with $L_{\text{tot}}$ as the independent variable suggest a strong dependence of short-term annoyance on $L_{\text{tot}}$ (see \autoref{tab:my-table}). The slopes equal 0.29 ($R^2=0.64$) for $T1$ and 0.30 ($R^2=0.62$) for $T2$. $\beta^{\ast}$ equals 0.79 for both regressions. 
Regarding dispersion, the interquartile ranges observed were always below 3.0 units on the annoyance scale.   

In addition, we computed linear regressions for each specific sound from $S3$ to $S6$. Looking at the linear regressions  (\autoref{fig:AnnVsEm1} and \autoref{fig:AnnVsLp1}), the specific sounds $S5$ - with ground effect - and $S6$  - without ground effect - appear to be slightly more annoying than $S3$. The ANOVA test showed that, indeed, the specific sounds had a statistically significant impact on the rating of short term annoyance, see \autoref{tab:ANOVA} ($F(2.1, 124.3)=7.2$, $p < .001$). However, the low generalized effect size of $0.005$ corresponding to the stimuli factor in the ANOVA \autoref{tab:ANOVA} should be taken into account. The influence of the type of specific sound on the annoyance is limited.

More specifically, a t-test made on the different wind turbine sounds revealed that the annoyance rating for $S6$ and $S5$ was significantly different from $S3$. This confirms the visual analysis of the linear regressions. It is also noteworthy that there is no statistical difference between $S5$ and $S6$, although the ground effect is not included in $S6$ whereas it is in $S5$.

 The impact of the type of specific sound on the annoyance rating increases with the emergence (\autoref{fig:AnnVsEm1}). This interaction between the emergence and the type of stimuli is confirmed by the ANOVA \autoref{tab:ANOVA} ($F(7.8, 451.5)=2.6$, $p = .01$).

\begin{figure*}[h]
\figline{\fig{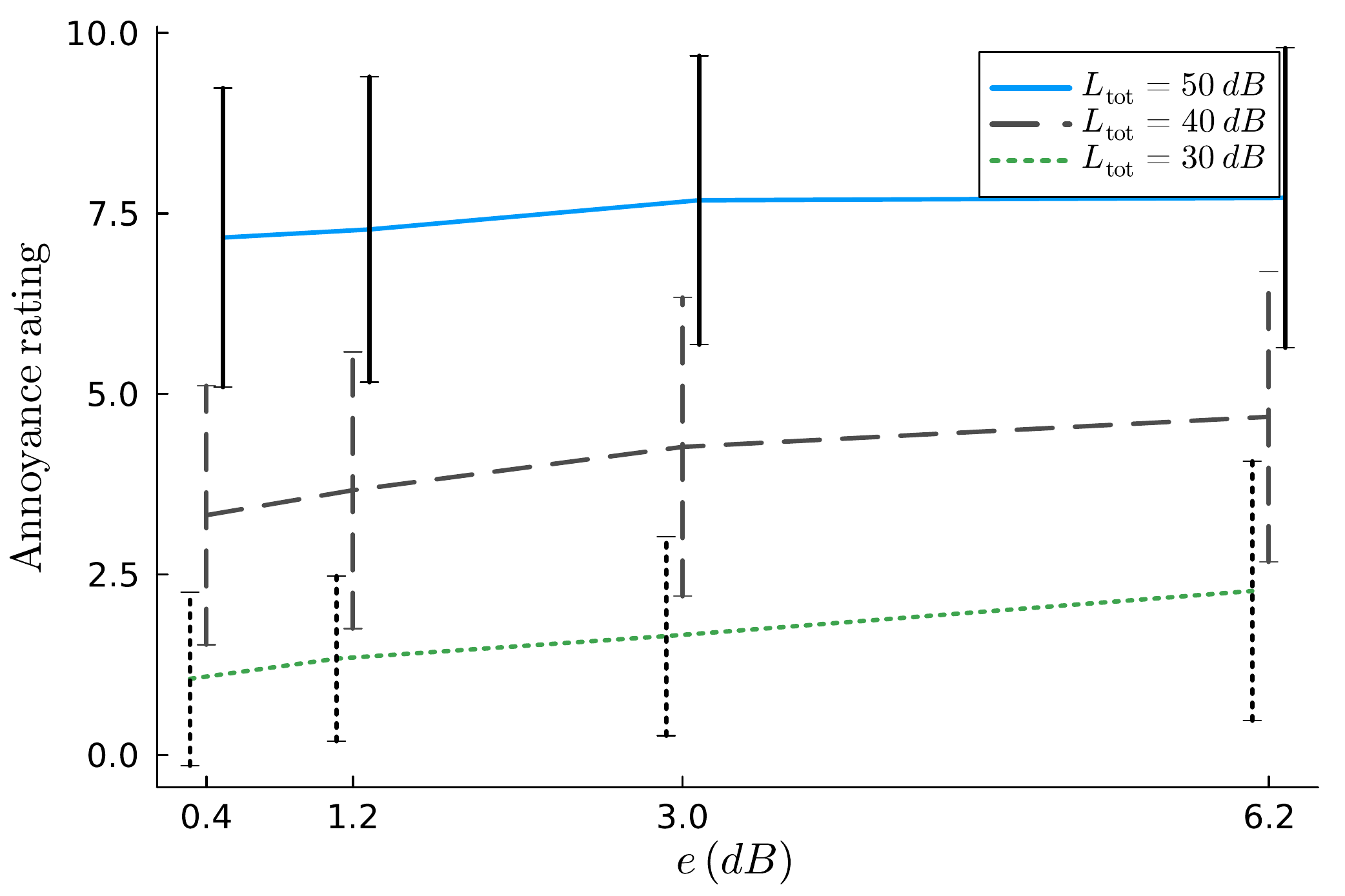}{8cm}{(a) T1 }\label{fig: mean level 1}
\fig{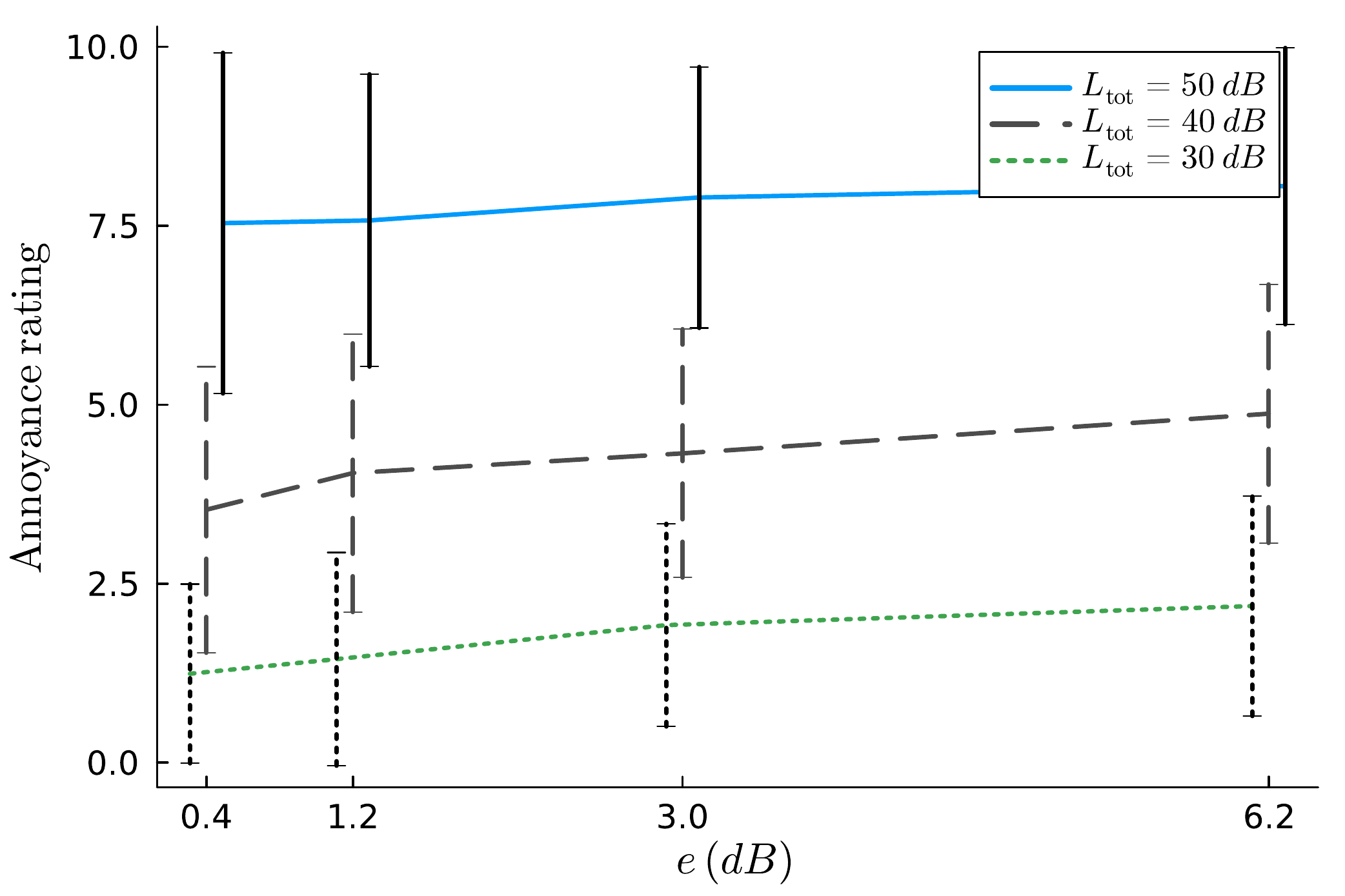}{8cm}{(b) T2 }\label{fig:mean level 2}}
\caption{Short-term annoyance ratings as a function of sound emergence for different sound pressure levels. 
For the sake of readability, the data series are slightly shifted horizontally to avoid overlap.}

\label{fig:mean_values1}
\end{figure*}

\autoref{fig:mean_values1} illustrates the relative dependence of the mean short-term annoyance on $e$ and $L_{\text{tot}}$. $L_{\text{tot}}$ has a greater effect on annoyance. The effects of variations in $e$ seems to have more impact on the annoyance at lower sound pressure levels. 
\autoref{fig:mean_values1} also illustrates the interaction between $e$ and $L_{\text{tot}}$ that are reported as statistically significant by the ANOVA test \autoref{tab:ANOVA} ($F(5.2, 302.6)=4.5$, $p < .001$): the lower the emergence the higher the influence of the sound pressure level.


Since the effects of $e$ and $L_{\text{tot}}$ on short-term annoyance are both statistically significant, we carried out a multi-linear regression with $e$ and $L_{\text{tot}}$ as the independent variables. However, compared to the linear regression where $L_{\text{tot}}$ is the only independent variable (RMSE=1.89 for $T1$), the two-variable model brings only marginal improvement in the prediction of short-term annoyance (RMSE=1.85 for $T1$). Similar variations in RMSE are observed for $T2$ (\autoref{tab:my-table}).

\begin{figure*}[h]
\figline{\fig{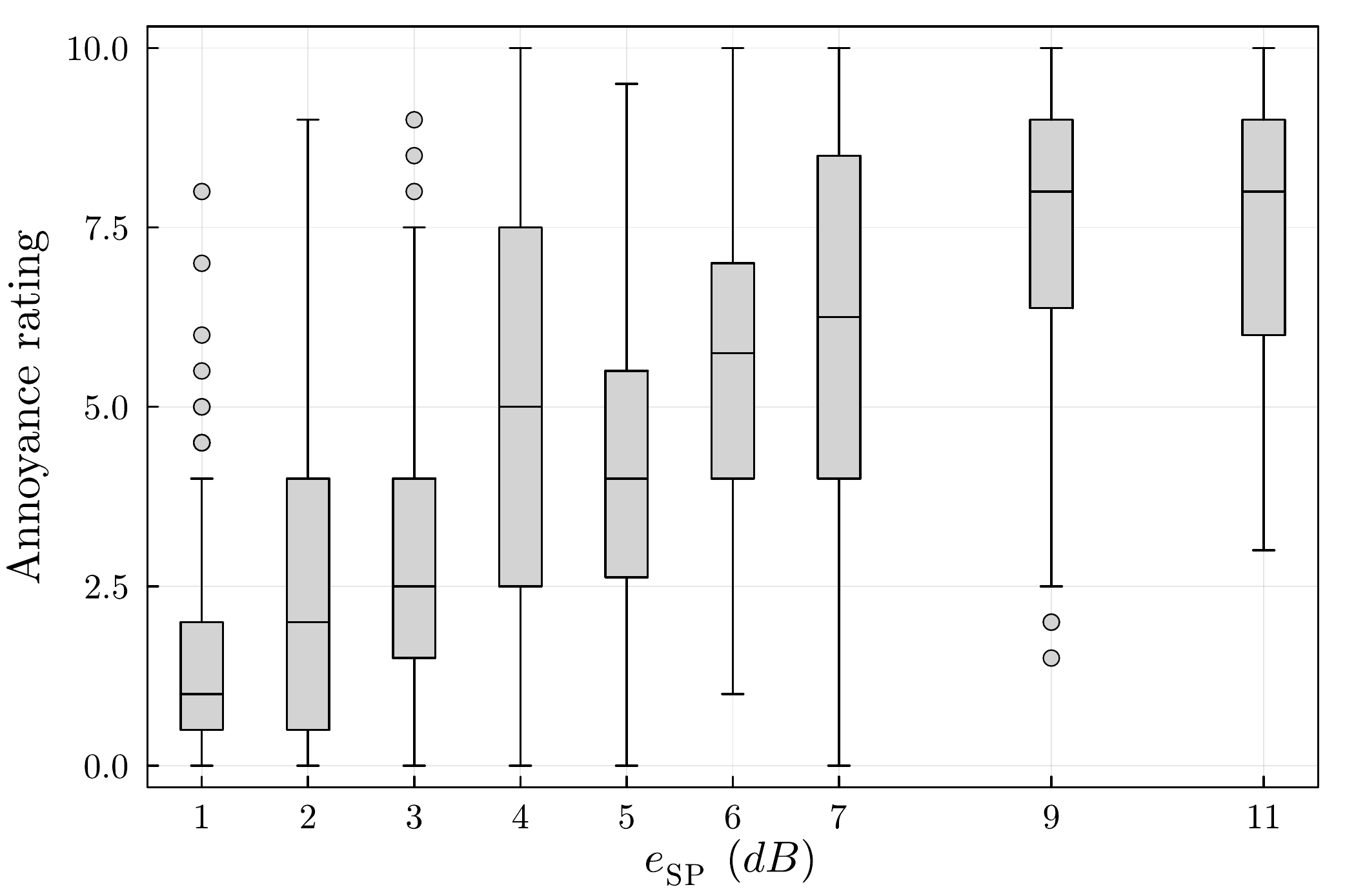}{8cm}{(a) T1}\label{fig:e_1000}
\fig{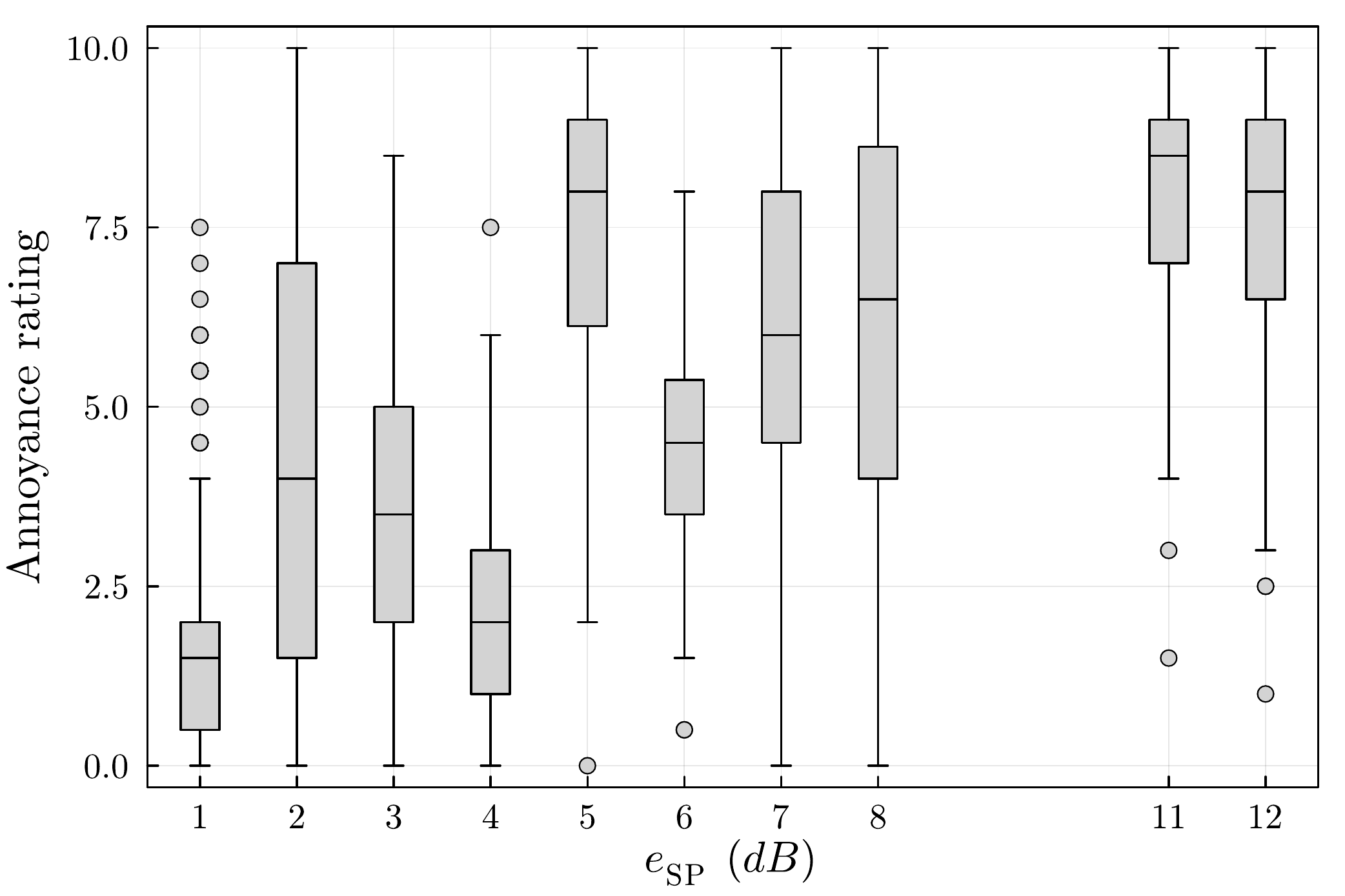}{8cm}{(b) T2}\label{fig:e_2000}}
\caption{Short-term annoyance rated as a function of spectral emergence in $T1$ and $T2$. The spectral emergence corresponds to the octave centered around 250 Hz, which is the octave with the maximum emergence in both tests.}
\label{fig:spectral emergence}
\end{figure*}

Switching now to the spectral emergence $e_{\text{SP}}$ obtained from octave-band equivalent sound pressure levels (Eq. \ref{eq:eSP}), an ANOVA test is not possible because the specific sounds have different $e_{\text{SP}}$. With a maximal difference of $3.7~dB$ between $S3$ and $S6$ at $e = 5~dB$ and $L_{\text{tot}} = 50~dB$ in $T2$), the design of the test is not complete for this variable. However, it is possible to compute linear regressions (see \autoref{tab:my-table}). There, it can be noticed that the standardized coefficients $\beta^{\ast}$ for $e_{\text{SP}}$ are even lower than for $e$.
This result is consistent with the box and whisker plots showed in \autoref{fig:spectral emergence}. It is clearly visible that, while there is a general trend in the variations of short-term annoyance as a function of $e_{\text{SP}}$, with very low values of $e_{\text{SP}}$ corresponding to a lower short-term annoyance and very high values of $e_{\text{SP}}$ corresponding to higher short-term annoyance, the correlation is weaker in the mid-range values of $e_{\text{SP}}$.  
The range of variation of $e_{\text{SP}}$ is 11~dB, which is about twice the range of $e$ in this study, whatever the test. By definition $e_{\text{SP}}$ is obtained from calculations of $e$ in octave bands from 125~Hz to 4~kHz. In our case, however, the 250~Hz octave band was always the band with the highest value of sound emergence, whatever the tuple $(S_x,R_y)$ analyzed.      

The box and whisker plots obtained with SRR as the independent variable are not reproduced here but they 
appear to be very similar to those with $e$ as the independent variable (\autoref{fig:AnnVsEm1}). The most notable difference is the lower slope ($\beta=0.07$, $R^2=0.02$ for $T1$)  when compared to $e$. However, the lower slope can be explained by the relatively larger range of variation of SRR with respect to $e$ and the standardized coefficients are the same ($\beta^{\ast} = 0.13$ for $T1$). Similar values were obtained for $T2$ (\autoref{tab:my-table}). 

When loudness metric is selected as the independent variable, the results of the same statistical analysis for the three indicators considered in this study are very similar (\autoref{tab:my-table}), so only those for $N$ will be given here. 
Like for $L_{\text{tot}}$, the data for both $T1$ and $T2$ is less scattered with $L_N$ around the medians than with $e$ \autoref{tab:my-table}.  
The linear regression returns slopes that are intermediate ($\beta =0.23$ for $T1$) between those obtained with $e$ and $L_{\text{tot}}$ as the independent variable, whereas the coefficients of determination ($R^2=0.61$ for $T1$) are as high as those obtained with $L_{\text{tot}}$ (\autoref{tab:my-table}). The standardized coefficients ($\beta^{\ast} = 0.78$ for $T1$ and $\beta^{\ast} = 0.79$ for $T2$ ) are slightly lower than the ones for $L_{\text{tot}}$.

\subsection{Audibility}
\begin{figure*}[h]
\figline{
\fig{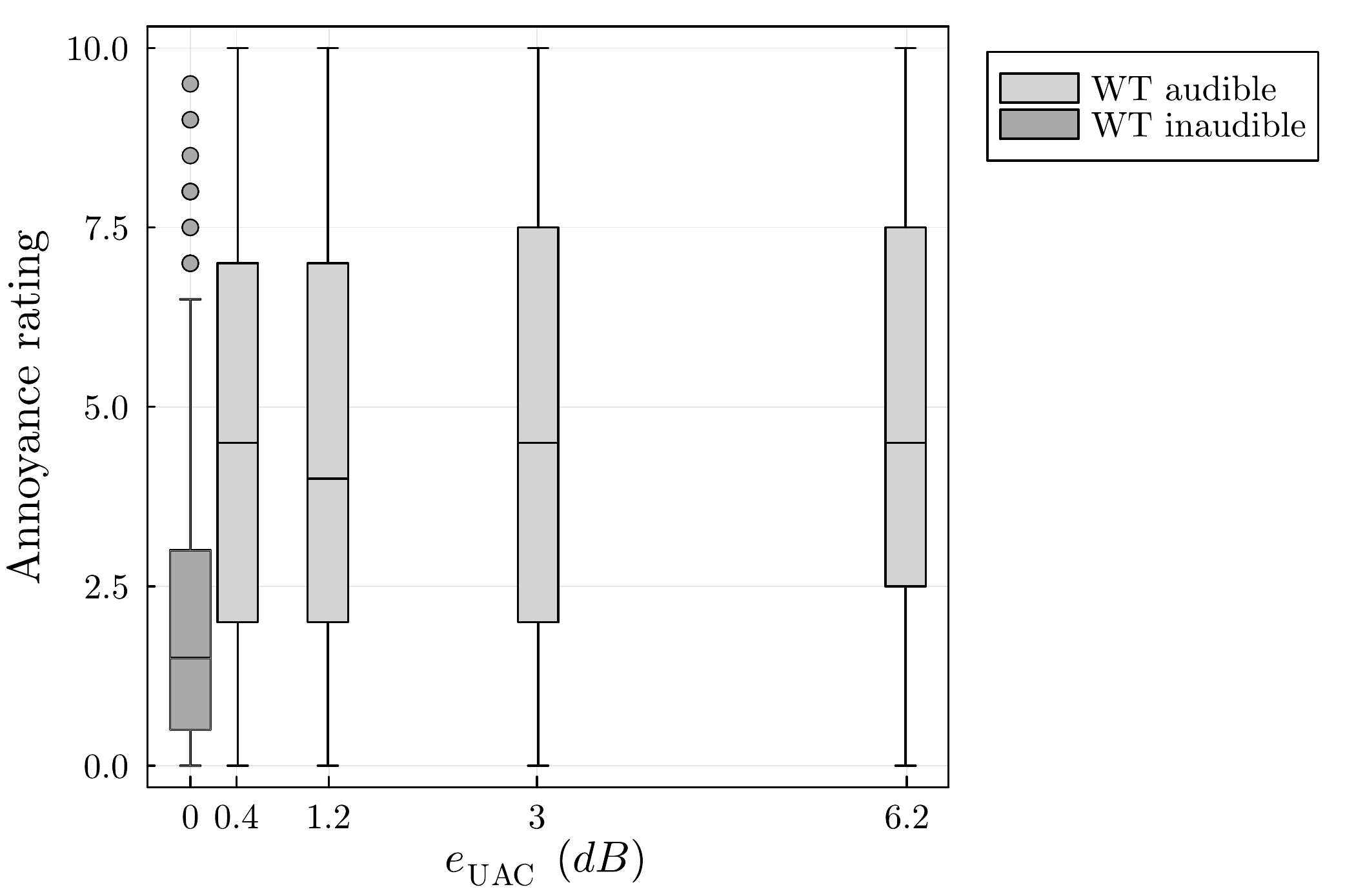}{9cm}{(a) T1}\label{fig:emergence 1 hyp}
\fig{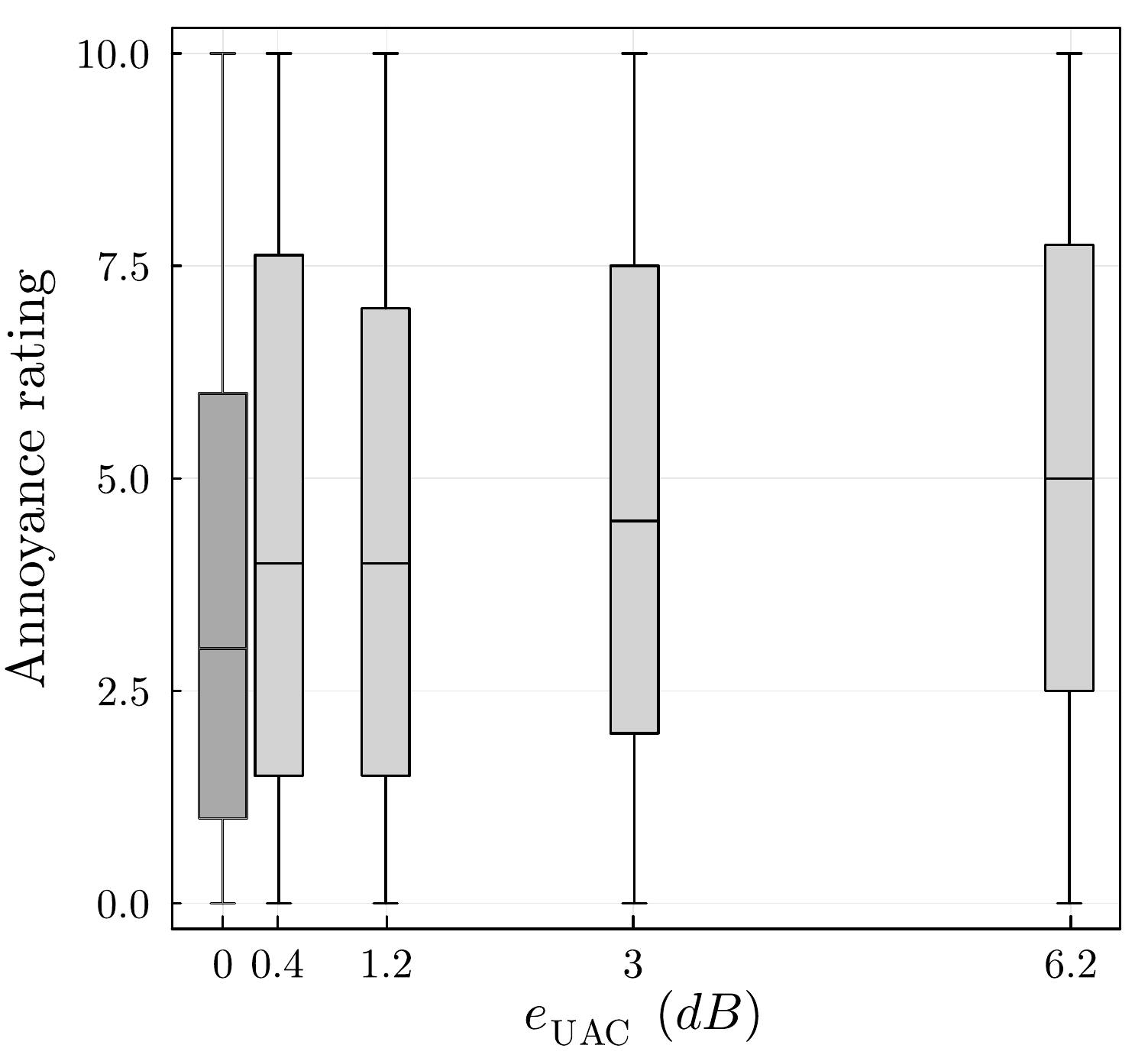}{6.75cm}{(b) T2}\label{fig:emergence 2 hyp}
}
\caption{Short-term annoyance ratings as a function of emergence under audibility condition.}
\label{fig:AnnVse_hyp}
\end{figure*}

Based on the answers to the question "Can you hear the wind turbine?" we also tested the correlation between short-term annoyance and sound emergence under audibility condition $e_{\text{UAC}}$ (see \autoref{eq:eUAC}). Each test represents 1440 answers in total. The subjects heard a wind turbine 1191 times in $T1$ and 1168 times in $T2$.  In \autoref{fig:AnnVse_hyp}, it can be seen that short-term annoyance does not systematically increase with $e_{\text{UAC}}$  when the specific sound is audible, while the annoyance rating is much lower when the subjects do not detect the wind turbines. 

\begin{figure*}[h]
\ifthenelse{\boolean{color}}
{\figline{\fig{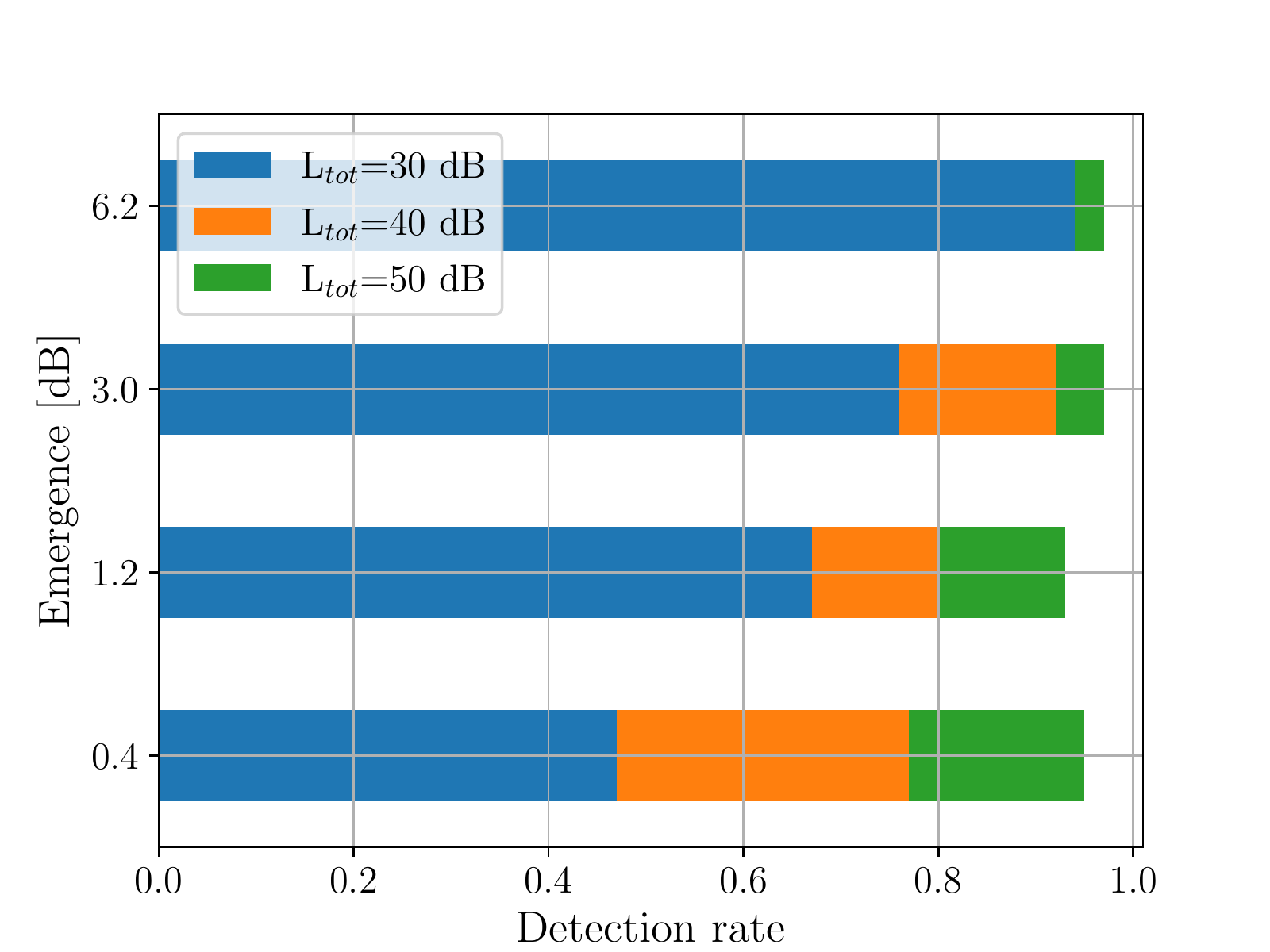}{8.5cm}{(a) T1}\label{fig:detection 1}
\fig{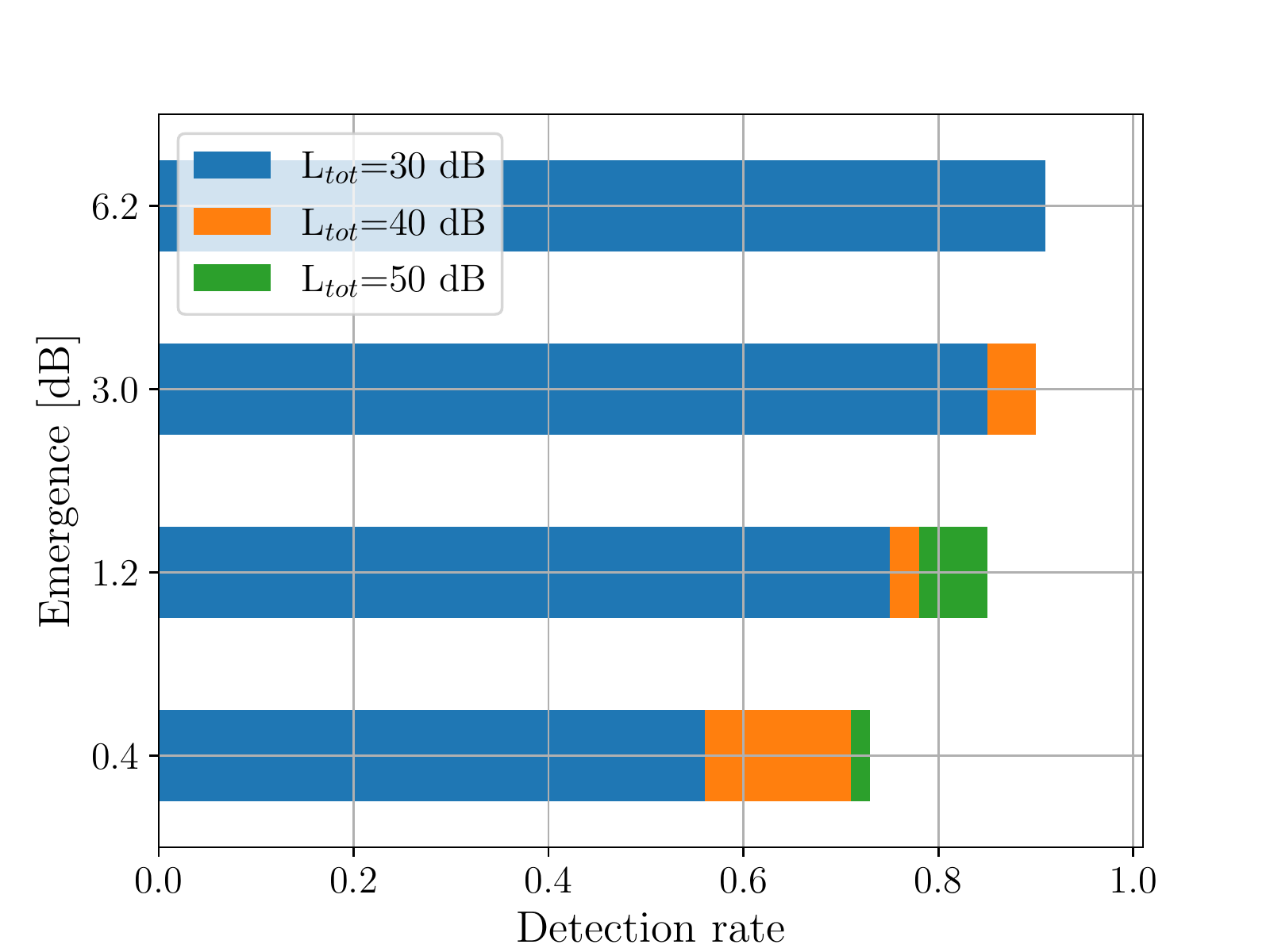}{8.5cm}{(b) T2}}\label{fig:detection 2}}
{\figline{\fig{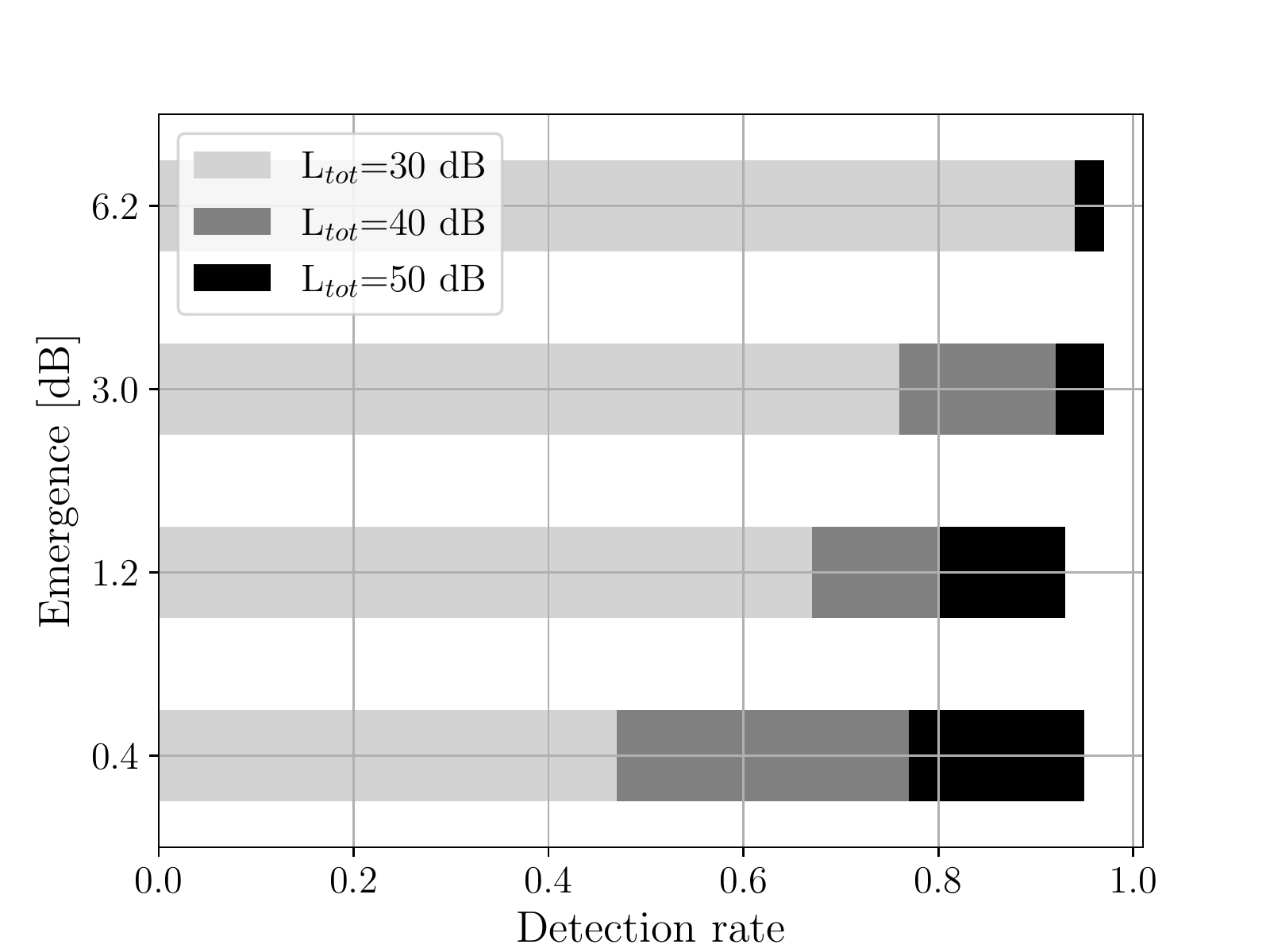}{8.5cm}{(a) T1}\label{fig:detection 1}
\fig{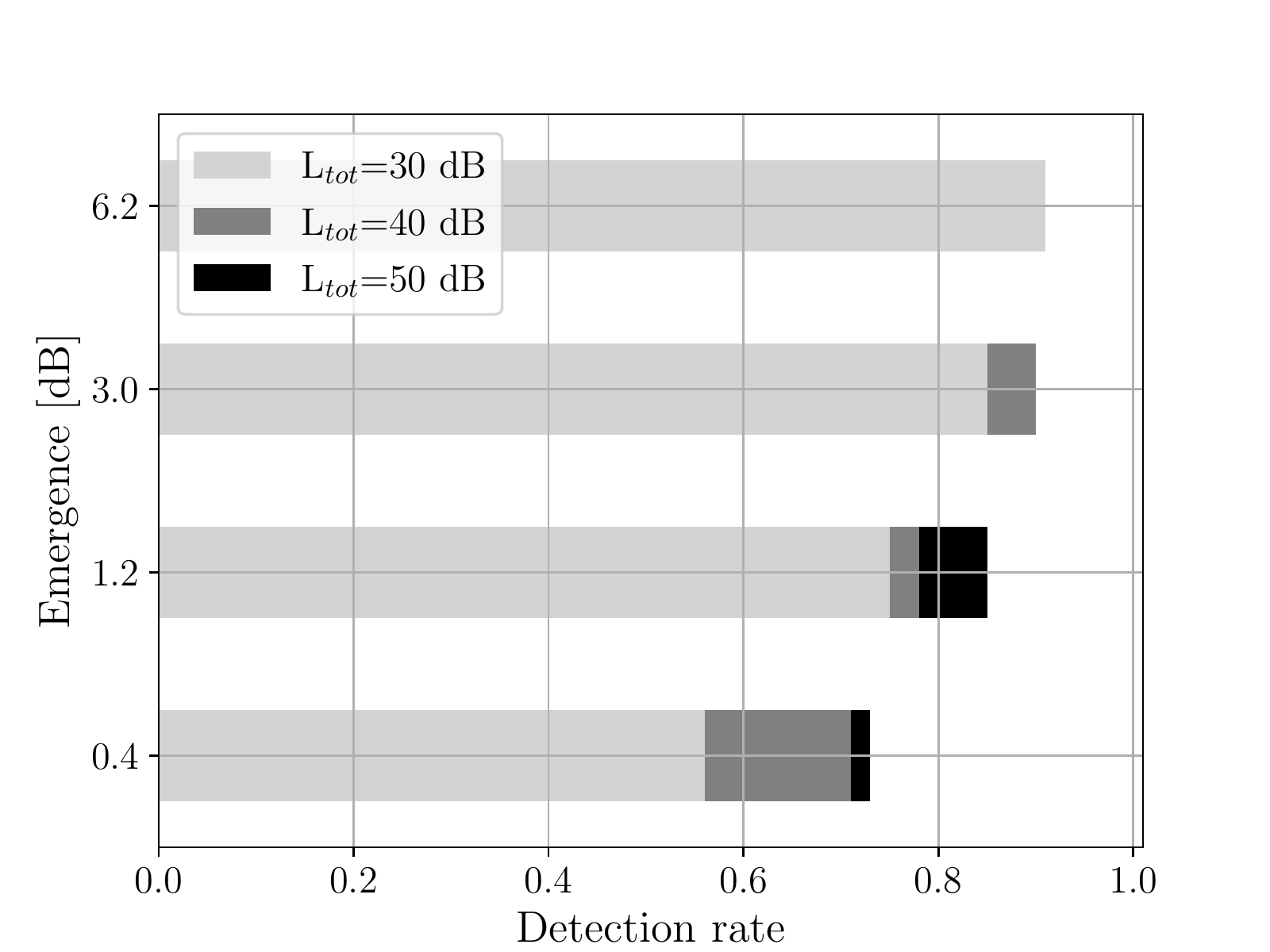}{8.5cm}{(b) T2}}\label{fig:detection 2}}
\caption{Audibility rates, as a function of the emergence and of $L_{\text{tot}}$. In this figure, the bars are not stacked but superimposed. For a specified value of $e$, the bar representing the audibility rate for $L_{\text{tot}}=50$~dB is first displayed, before the bar for $L_{\text{tot}}=40$~dB is added. The bar for $L_{\text{tot}}=30$~dB is displayed last. In other words, the audibility rate for a specific $L_{\text{tot}}$ can be read at the right end of the corresponding rectangular area.}
\label{audibility1}
\end{figure*}

In addition to allowing for the calculation of $e_{\text{UAC}}$, the answers to the question "Can you hear a wind turbine?" enable to calculate audibility rates for the different sound stimuli presented. As expected, wind turbine sounds are more easily detectable as sound emergence increases (\autoref{audibility1}). But the subjects' ability to detect a wind turbine sound depends also strongly on $L_{\text{tot}}$ for the lower values of $e$ in the case of $T1$ (\autoref{audibility1}). While the global trend is that the variation of the empirical audibility rate is an increasing function of $L_{\text{tot}}$, the dependence was not perfectly monotonous for $e=3.0$ and $e=6.2$~dB. 
Moreover, even for the lowest sound emergence $e=0.4$~dB and the lowest $L_{\text{tot}}$ considered in our study, the audibility rate is quite high at 47~\% for $T1$ and 56~\% for $T2$. With respect to the averaged audibility rates for all the stimuli, we found a 5~\% higher rate for stimuli that were based on $S5$ and a 7~\% higher one for those that were based on $S6$.   

\section{Discussion}
\label{sec:discuss}

\subsection{Short-term annoyance}
The main finding of our study is that the $L_{\text{tot}}$ predicts short-term annoyance due to wind turbine sounds much better than $e$ does. 
Even if the correlation between short-term annoyance and $e$ is statistically significant, $e$ appears as a second-order descriptor. The linear regressions between short-term annoyance and the two indicators show that, in both $T1$ and $T2$, the $R^2$ coefficients and the slopes are higher for the $L_{\text{tot}}$ than for $e$ while the $RMSE$ are lower. In addition, $\beta^{\ast}$ is six times higher for $L_{\text{tot}}$. The answers of the subjects are also less scattered in the regressions between $L_{\text{tot}}$ and annoyance  (\autoref{fig:AnnVsLp1}) than in the regressions between $e$ and annoyance (\autoref{fig:AnnVsEm1}). Finally, the ANOVA test reveals a better correlation between short-term annoyance and $L_{\text{tot}}$ as well, with a higher $F$ value than the one corresponding to $e$. $L_{\text{tot}}$ also explains much more variance of the annoyance rating than $e$ with respective generalized effect size of $0.65$ and $0.04$.

Therefore, the link between $L_{\text{tot}}$ and short-term annoyance is stronger than the one between $e$ and short-term annoyance. This is consistent with other recent experiments dealing with wind turbine sounds \cite{Schaffer:2016ww} that found that short-term annoyance is strongly linked to $L_{\text{tot}}$ as well. More generally, the relatively minor contribution of the residual sound to the annoyance was also emphasized by a wide meta-analysis of socio-acoustic surveys about annoyance from environmental noise, although the research results analysed did not cover noise from wind energy \cite{Fields:1998wv}.

Furthermore, a linear model where short-term annoyance is predicted by $L_{\text{tot}}$ alone is not significantly improved by adding $e$ as a second independent variable. Therefore, if one is to use a single descriptor, $L_{\text{tot}}$ should be preferred among these two independent variables. 

Using the linear model with $L_{\text{tot}}$ as the only independent variable (\autoref{tab:my-table}), at $L_{\text{tot}}=35$~dB, the annoyance ratings of 2.8-2.9 on the ISO 15666 11-point scale \cite{ISO:2003aa} are similar to other laboratory tests in \cite{Schaffer:2016ww}. Extrapolating the same model to $L_{\text{tot}}=55$~dB leads to annoyance ratings of 8.6-8.9, which is also similar to previous research \cite{Schaffer:2016ww}. At constant $L_{\text{tot}}$, only a minor effect of the specific sound on short-term annoyance for $S3$ to $S6$ was found. This is in accordance with \cite{Dutilleux:2020aa} where there was little difference in the annoyance rating between the three specific noises considered.     

Compared to $T0$ \cite{Dutilleux:2020aa}, the results from $T1$ and $T2$ about the relative merits of sound emergence and sound pressure level are almost identical, although (a) another set of stimuli based on different turbine types from different wind parks was used to build the library of sound stimuli, (b) the range of $L_{\text{tot}}$ considered was not identical, (c) the sound reproduction system used was different, with headphones instead of loudspeakers, (d) $T1$ and $T2$ share no subject with $T0$.  While $R2$ and $R0$ present some similarities, the nature of $R1$ is different because it is dominated by wind-induced noise from vegetation. In addition, $S0$ to $S2$ were obtained at fairly low wind speed at hub height, whereas $S3$ to $S5$ correspond to more than twice higher wind speeds. 

While \cite{Dutilleux:2020aa} considered $e$ as the only possibility to define sound emergence, two variants of $e$, namely $e_{\text{UAC}}$ and $e_{\text{SP}}$ were evaluated in the present study. The correlation between annoyance and $e_{\text{UAC}}$ increased slightly when audibility was taken into account. But the relationship between $e_{\text{UAC}}$ and short-term annoyance appeared to be non-linear. It was mostly under the control of audibility with relatively little influence of the SRR, as most of the variations of short-term annoyance occurred close to $e_{\text{UAC}}=0$ dB.  In addition, $e_{\text{UAC}}$ would be more difficult to use in practice because of the requirement to decide on audibility. 
When $e_{\text{SP}}$ was used, the correlation with short-term annoyance was lower, although the range of variation of $e_{\text{SP}}$ was larger than that of $e$. This suggests that eliminating $e_{\text{SP}}$ from the French legislation on wind turbine noise was a harmless decision. 

The poor correlation between short-term annoyance and the different forms of emergence considered in this study, and likewise between short-term annoyance and SRR, suggests that relying on indicators that depend on $L_{\text{res}}$ is not necessarily beneficial compared to indicators based solely on $L_{\text{tot}}$. 

In the search of alternative one-variable linear models to better predict short-term annoyance, loudness level $L_N$ and its two variants $L_{N_{max}}$ and $L_{N_5}$ seemed like relevant candidates because of their potential to get closer to a \emph{perceived} $L_{\text{tot}}$. Our finding, however, is that they do not outperform $L_{\text{tot}}$. Bearing in mind that the range of sound pressure levels investigated here is close to the 40-phones equal loudness curve that was used to define A-weighting, it should not come as a surprise that $L_N$ and $L_{\text{tot}}$ exhibit similar performances. Since $L_N$ is more complex to calculate \cite{ISO:2017ww} than $L_{Aeq}$, using $L_N$ to predict annoyance is not justified in the present case.

To summarize the results of our analysis of the potential of different indicators to predict short-term annoyance, the most obvious replacement to $e$ for assessing annoyance due to wind turbine noise is $L_{\text{tot}}$ estimated by the A-weighted equivalent sound pressure level, despite the well-known limitations of the latter indicator in general and in the specific case of wind energy \cite{Persson-Waye:2002tm, Haac:2019uv}. 

The difficulty here is that the causes for annoyance due to wind turbine sounds cannot be accounted for by a one-variable model. It is well known that other factors than $L_{\text{tot}}$ such as tonal components \cite{oliva_annoyance_2017, yokoyama_perception_2016} or periodic sound pressure level fluctuation - often called  "amplitude modulation" for the sake of conciseness -  \cite{Schaffer:2018ub,Pohl:2018aa} play a role in short-term annoyance. Other authors have pointed out the relatively minor contribution of amplitude modulation \cite{Schaffer:2016ww}, whereas a larger spread in annoyance scores was found in earlier research \cite{Persson-Waye:2002tm}, depending on the sound stimuli. The latter publication, however, considered submegawatt-scale turbines and both two- and three-bladed turbines. The present study found an effect of the spectro-temporal characteristics of the specific sounds on short-term annoyance that is much lower than that of $L_{\text{tot}}$. These other factors could come as a correction applied to an estimator of $L_{\text{tot}}$ to define a \emph{rating level} \cite{ISO:2016aa}. This is likely to improve the prediction of short-term annoyance. But, this goes beyond the scope of our study.\\

\subsection{Audibility}

Since $e$ does not reflect the intensity of the noise exposure, a possible interpretation for the use of $e$ when setting limit values in legislation is that $e$ would reflect the audibility of a sound source in the environment. However, in both $T1$ and $T2$ a large part of the subjects were able to hear the wind turbine in situations of negative SRR which correspond to $e$ as low as 0.4 dB. These results are in line with the findings of \cite{Dutilleux:2020aa} and they raise the issue of the psycho-acoustical relevance of the 3~dB or 5~dB sound emergence thresholds found in the existing national legislation and international guidelines that rely on this indicator to set noise limits. This issue was highlighted in \cite{Viollon:2004aa} and is further discussed in \cite{Dutilleux:2020aa}.

There are obvious differences between the audibility rates found in $T1$ and $T2$, especially for the two lower values of $e$ considered. This reflects the influence of the residual sound as a potential masker. At constant $e$, $R1$ appeared to be masking the wind turbine sounds more efficiently than $R2$, probably because the power spectrum density of $R1$ was flatter than $R2$ in the frequency range where human hearing is most sensitive. As mentioned earlier, $R1$ is dominated by noise from foliage in the wind. Qualitatively this concurs with existing evidence that vegetation noise has a potential to mask wind turbine sounds \cite{Bolin:2010aa}. In addition, $R1$ was more steady than $R2$ with fewer instants of relative silence. 

The higher audibility rates found for $S5$ and $S6$ suggest that a source-specific parameter such as amplitude modulation increases is associated with higher detection rates. Amplitude modulation is clearly visible in \autoref{fig:S5}. Again, this is in line with the existing literature.

The audibility rates obtained also suggest that the influence of sound emergence on detection is more important at lower sound pressure levels than at higher ones. Whereas $e$ appears to correlate with the detectability of the wind turbine for $L_{\text{tot}}=30$~dB, for $L_{\text{tot}}=50$~dB the subjects almost systematically detected the wind turbine regardless of $e$. Similarly, there is a correlation between $L_{\text{tot}}$ and the detectability of the wind turbine at low $e$ but the influence of $L_{\text{tot}}$ becomes negligible for a sound emergence as high as 6~dB. For $e=6$~dB, the wind turbine sound was indeed detected almost all the time by all the subjects. In this regard, while we found sound pressure level to be a strong predictor of audibility, like in previous research \cite{Haac:2019uv}, in our study sound emergence and sound pressure level seemed to complement each other to a certain degree when it comes to predicting the audibility of wind turbine sound. 

Comparing the audibility rates (\autoref{fig:detection 2}) with short-term annoyance ratings as a function of $e$ (\autoref{fig:AnnVsEm1}) and $L_{\text{tot}}$ (\autoref{fig:AnnVsLp1}) suggests that there is no clear link between audibility and short-term annoyance. Indeed other authors have stated they are affected by distinct factors \cite{Haac:2019uv}. In both $T1$ and $T2$, high audibility rates were observed at $e=6.2$~dB and $L_{\text{tot}}=30$~dB, while the average annoyance was low. Again, this raises the issue of the relevance of the 3 or 5 dB thresholds found in existing legislation.

\subsection{Methodological aspects}

Our study assessed only short-term annoyance as the stimuli lasted for only 20~s for a total duration of the listening test of 25 minutes to keep the subjects focused \cite{Nordtest:2002tr}, whereas in real life the exposure to wind turbine sounds is long-term. Nevertheless, the sound stimulus duration that was long enough to record several rotations of the turbines under investigation. Moreover, it appears to be well within the range of values found in the literature for listening tests relating to perception of wind turbine sounds \cite{Alamir:2019wf}. 

Although the subjects have been asked to project themselves into a long term situation when rating the annoyance, one cannot ascertain that their behavior would be identical after a long exposure to a wind farm. A laboratory study like ours cannot completely replace a field study. Such a field study would, however, raise serious difficulties due to the need to measure sound emergence in receptor position where the signal-to-noise ratio would be typically very low \cite{Dutilleux:2019aa}, leading to large uncertainties. Relying on noise prediction instead of measurement would not be an option either, because the uncertainty of the prediction would be too large to predict $e$ \cite{Dutilleux:2019aa}. 

In this experiment, headphones were used for sound reproduction. During the calibration, frequencies below 100 Hz were not considered. Therefore it is not possible to assess the fidelity of the reproduction of the sound stimuli that were acquired with a recording chain that features a lower cut-off frequency of 20 Hz. But headphones are not ideally suited to the reproduction in the lower end of the audio spectrum. Both the specific sounds and the residual sounds contain a lot of energy below 100 Hz (\autoref{fig:fft}). This part of the spectrum may contribute to the overall annoyance, and wind turbine sounds are known to be rich in low frequencies in the 20-200~Hz range \cite{Moller:2011wi}.   

Furthermore, limiting the duration of the listening test leads also to restricting the number of sound stimuli proposed in our listening tests, and one can wonder what can be inferred from a small set of controlled stimuli. Nevertheless, when the present study and \cite{Dutilleux:2020aa} are combined, they represent a variety of 5 modern horizontal axis turbines from 4 different wind farms. Moreover, these turbines were presented to the subjects in three different acoustic backgrounds. 

Further variability with respect to the previous study \cite{Dutilleux:2020aa} was introduced in this study by an attempt to account for the spectral modifications caused by ground effect. 
A limitation here is that a fixed sound propagation was simulated. In reality, sound propagation is subject to fluctuation, owing to atmospheric turbulence and to the cyclical variation of the distribution of source heights above the ground, as the turbine blades are rotating. Accounting for this is likely to increase the audibility rates of the wind turbines everything else being constant.

A possible shortcoming in our evaluation of $e_{\text{SP}}$ as a predictor of short-term annoyance is that all the combinations of specific and residual sounds considered in the two listening test led to a value of $e_{\text{SP}}$ obtained from the 250 Hz octave band. The correlation between $e_{\text{SP}}$ and annoyance may be stronger if higher octave bands had a bearing on $e_{\text{SP}}$. 

The listening tests this paper is based on were mainly designed to study annoyance and not the detection of wind turbine sound. A more in-depth analysis would be necessary to further apprehend this subject. It would also be interesting to add a distracting activity during a listening test and study the noticeability of the wind turbines. A potential limitation of our study is the absence of control stimuli, \textit{i.e.} stimuli without wind turbine sound.

\section{Conclusion}
\label{sec:conclusion}

From recordings of wind turbine sounds and residual sounds, we constructed two sets of synthetic soundscapes where the control parameters are the total sound pressure level and the specific-to-residual ratio. These soundscapes, that could have been recorded in many different places in Europe, formed the basis of two listening tests that account for situations with and without wind close to a hypothetical receptor location.

In the light of the results of the listening tests, (1) standard sound emergence $e$ has no major influence on short-term annoyance, although the correlation between standard sound emergence and short-term annoyance is statistically significant; (2) with respect to $e$ the two alternative definitions of sound emergence $e_{\text{UAC}}$ and $e_{\text{SP}}$ evaluated here do not bring any clear improvement, if any, to the correlation between sound emergence and short-term annoyance; (3) short-term annoyance from wind turbine sound is much better predicted by the total sound pressure level than by sound emergence. These conclusions hold when both, $e$, $e_{\text{UAC}}$ and $e_{\text{SP}}$ are based on equivalent sound pressure levels. Moreover, replacing the sound pressure level by loudness did not improve the correlation with short-term annoyance. 

It seems that the standard sound emergence based on equivalent sound pressure levels fails to account for the contribution of the residual sound to the annoyance from wind turbine sounds, and that an alternative metric must be developed to account for the role of the residual sound in annoyance that was observed by several authors. 

With the sound stimuli used in our listening test, it was also observed that sound emergence was a poor predictor of the audibility of wind turbine sounds. Indeed, even for values of sound emergence that are close to the lower bound of this indicator, more than 4 subjects out of 10 could detect the wind turbine sound in the various stimuli proposed. 

Sound emergence has been introduced in a number of legal texts and guidelines in Europe and at the international level apparently without any systematic evaluation of the merits of this metric. The results obtained here concur with those from a preliminary study - which were obtained with different stimuli, with a different sound reproduction method and with different subjects - to question the relevance of sound emergence when it comes to assessing the environmental impact of wind turbines. The two studies combined represent more than 90 subjects, 3 different residual sounds covering various situations and 6 different specific sounds from state-of-the-art wind turbines.

To the authors' knowledge, the regulations based on the reference to the residual sound pressure level and the specific-to-residual ratio do not rely on solid scientific evidence either. Our study suggests that the correlation between this indicator and short-term annoyance is also quite weak, which warrants further research.

\begin{acknowledgments}
The authors would like to thank Céline Angonin, MSc student in acoustics at NTNU who participated in field measurements at Storheia and Bessakerfjellet ; TrønderEnergi AS for sharing wind data ;  Tim Cato Netland, NTNU, for his help during the calibration experiments ; U. Peter Svensson, NTNU, for his feedback on the manuscript ; and the 60 volunteers who took the listening tests. 
\end{acknowledgments}






\bibliography{papers}




\end{document}